\documentclass[journal]{IEEEtran}
\IEEEoverridecommandlockouts

\usepackage[utf8]{inputenc}
\usepackage{makeidx}
\makeindex
\usepackage[T1]{fontenc}
\usepackage[pdftex]{graphicx}
\usepackage{cite}
\usepackage{url}
\usepackage{amsmath}
\usepackage{amssymb}
\usepackage{multirow}
\usepackage{eso-pic}
\usepackage{balance}
\usepackage{listings}
\usepackage{xspace}
\usepackage{subfigure}

\pdfoutput=1
\usepackage{hyperref}
\hypersetup{
    colorlinks=true,
    linkcolor=black,
    citecolor=black,
    filecolor=black,
    urlcolor=black,
}

\usepackage[printonlyused]{acronym}

\usepackage{color}

\newcommand{\etal}{et al.}
\newcommand{\sume}[0]{NetFPGA SUME }
\newcommand{\ewedge}[0]{Edgecore Wedge 100BF-32X }

\usepackage[]{algorithm2e}
\DontPrintSemicolon
\SetAlCapSkip{1em}
\SetKwInOut{Input}{input}
\SetKwInOut{Output}{output}
\SetKwFunction{Print}{print}
\SetKwFunction{AllPaths}{allPaths}
\SetKwFunction{End}{end}

\usepackage{siunitx}
\DeclareSIUnit{\Bit}{Bit}
\DeclareSIUnit{\Byte}{Byte}

\newlength{\figsize}
\newlength{\subfigwidth}
\newlength{\subfiglabelwidth}
\newcommand\fig[1]{Figure~\ref{fig:#1}}

\newcommand\sect[1]{Section~\ref{sec:#1}}

\newboolean{makevspace}
\newcommand{\cvspace}[1]{%
    \ifthenelse
        {\boolean{makevspace}}
        {\vspace{#1}}
        {}%
    }
\begin{document}
  
\title{P4-IPsec: Site-to-Site and Host-to-Site VPN with IPsec in P4-Based SDN}

\author{Frederik~Hauser,~
        Marco~Häberle,~
        Mark~Schmidt,~
        Michael~Menth%
\thanks{Frederik Hauser, Marco~Häberle,
        Mark~Schmidt,
        and Michael~Menth are with Chair~of~Communication~Networks, University~of~Tuebingen, Tuebingen, Germany. E-mail: \{%
       frederik.hauser,%
       marco.haeberle,%
       mark-thomas.schmidt,%
       menth\}@uni-tuebingen.de}%
\thanks{This work was supported by the Deutsche Forschungsgemeinschaft (DFG) under grant ME2727/1-2 and the bwNET100G+ project which is funded by the Ministry of Science, Research and the Arts Baden-Württemberg (MWK). The authors alone are responsible for the content of this paper.}}

\maketitle
  
\pagenumbering{gobble}
  
\begin{abstract}
In this work, we present P4-IPsec, a concept for IPsec in software-defined networks (SDN) using P4 programmable data planes.
The prototype implementation features ESP in tunnel mode and supports different cipher suites.
P4-capable switches are programmed to serve as IPsec tunnel endpoints.
We also provide a client agent to configure tunnel endpoints on Linux hosts so that site-to-site and host-to-site application scenarios can be supported which are the base for virtual private networks (VPNs).
While traditional VPNs require complex key exchange protocols like IKE to set up and renew tunnel endpoints, P4-IPsec benefits from an SDN controller to accomplish these tasks.
One goal of this experimental work is to investigate how well P4-IPsec can be implemented on existing P4 switches. 
We present a prototype for the BMv2 P4 software switch, evaluate its performance, and publish its source code on GitHub \cite{p4-ipsec-github}.
We explain why we could not provide a useful implementation with the NetFPGA SUME board.
For the Edgecore Wedge 100BF-32X Tofino-based switch, we presented two prototype implementations to cope with a missing crypto unit.
As another contribution of this paper, we provide technological background of P4 and IPsec and give a comprehensive review of security applications in P4, IPsec in SDN, and IPsec data plane implementations. 
According to our knowledge, P4-IPsec is the first implementation of IPsec for P4-based SDN.    
\end{abstract}
  
  
\IEEEpeerreviewmaketitle
  
\graphicspath{{figures/}}
  
\section{Introduction}
\label{sec:introduction}

\acfp{VPN} extend private networks across public networks by adding authentication and encryption to network traffic.
\ac{IPsec} is one of the oldest, but still most-widespread \ac{VPN} protocols.
Standardized by the IETF, it introduces protection on the \ac{IP} layer.
Due to its large distribution, many implementations for network appliances and operating systems are available.
Although it is criticized for its complexity, proven deployment patterns allow efficient and reliable operation.

IPsec tunnel setup requires user configuration plus keying material that is exchanged by IPsec peers via the \acf{IKE} protocol.
Complexity grows with the number of IPsec peers, especially in highly dynamic environments such as campus or enterprise networks with many users and sites.
Several works investigate on how to leverage the centralized control plane of \acf{SDN} to simplify \ac{IPsec} operation.
However, the possibilities for \ac{IPsec} deployment in \ac{SDN} were limited.
Typical \ac{SDN} switches have a fixed-function data plane that does not provide support for \ac{IPsec}.
As a result, \ac{IPsec} data plane processing needs to be moved to an additional software-based \ac{PPF}.
Besides being an additional component, this adds latency as traffic needs to be forwarded back and forth.
Programmable data planes as offered by P4 are a game changer.
Data plane behavior can be described in a high-level programming language.
Those network programs can be executed by software or hardware devices. 
For IPsec this means that instead of shifting \ac{IPsec} functionality to \acp{PPF}, functions such as \ac{IPsec} can be implemented directly on the data plane of \ac{SDN} switches.
In our previous work P4-MACsec \cite{p4-macsec}, we introduced MACsec for P4-based SDN.
We proposed a data plane implementation in P4 and introduced a novel concept for automated deployment and operation of MACsec.

In this paper, we present the first integration of \ac{IPsec} \ac{VPN} for P4-based \ac{SDN}.
We give an introduction on the technological background and provide an extensive survey on related work in that field.
We present an IPsec data plane implementation that integrates \ac{IPsec} components and processes with constructs and components under the given constraints of the P4 data plane programming language.
Cryptographic operations for authentication, encryption, and decryption are implemented in P4 externs where IPsec components such as the \ac{SPD} and \ac{SAD} are part of the P4 processing pipeline.
P4 switches that implement the functionality of P4-IPsec can be deployed in host-to-site and site-to-site VPN scenarios.
Control plane functions for IPsec operation are part of a central SDN controller that maintains IPsec tunnels without the help of distributed key exchange protocols such as \ac{IKE}.
As these components are steered by a centralized control plane through an authenticated and encrypted control connection, complex IKE-based key exchange protocols are substituted by controller-based tunnel setup and renewal procedures.
For host-to-site operation, we introduce a client agent for Linux operating systems that runs on the roadwarrior hosts.
It establishes an interface to the central \ac{SDN} controller via a gRPC connection.
To investigate how well P4-IPsec can be implemented on existing P4 targets, we work on three prototypes. 
We successfuly implement a prototype for the \ac{BMv2} P4 software target and conduct a performance evaluation.
We release the source code of our prototype along its testbed environment under the Apache v2 license on GitHub \cite{p4-ipsec-github}.
In addition, we report on implementation experiences for the \sume board and \ewedge P4 switch.
For the latter, we present two workaround implementation and compare them in performance experiments.

P4-IPsec introduces several benefits over traditional IPsec operation.
First, we improve scalability by making switches and roadwarrior hosts stateless components whose functionality is only managed by an SDN controller.
Second, we improve flexibility by converting P4 targets into IPsec endpoints, i.e., IPsec tunnels can terminate close to the network hosts that should be made accessible via the VPN.
This limits the size of the perimeter and improves security through better isolation.
Last, we encourage open networking research and operation.
Network functionality can be modified in agile development processes, source code can be audited and improved by a larger audience.

The rest of the paper is organized as follows.
\sect{foundations} gives an overview on \ac{IPsec} \ac{VPN} and data plane programming with P4.
In \sect{related-work}, we describe related work on P4-based network security applications, \ac{IPsec} in \ac{SDN}, and \ac{IPsec} data plane implementations.
\sect{concept} presents the architecture of P4-IPsec.
In \sect{implementation-mininet}, we describe the prototypical implementation of P4-IPsec with Mininet and \ac{BMv2}.
\sect{evaluation-mininet} presents the performance evaluation of that prototype.
In \sect{implementation-hardware}, we report implementation experiences for the \sume and \ewedge P4 targets.
\sect{conclusion} concludes this work.
The appendices include a list of the acronyms used in the paper.
\section{Technical Background}
\label{sec:foundations}

We give an introduction to \ac{VPN} with \ac{IPsec} and data plane programming with P4.

\subsection{IPsec VPN}
\label{sec:ipsec-foundations}
\acf{IPsec} is a widespread \ac{VPN} protocol suite.
It applies authentication and encryption on the \ac{IP} in host-to-host, gateway-to-gateway, and host-to-gateway communication scenarios.
RFC 4301 \cite{rfc4301} is the latest version of its specification.

\subsubsection{Protocols}
\ac{IPsec} comprises the \acf{AH} and \acf{ESP} protocol.
\ac{AH} \cite{rfc4302} protects \ac{IP} packets by sender authentication and  packet integrity validation.
It applies a hash function with a shared key (e.g., HMAC-SHA256) to calculate \acp{ICV} and adds packet sequence numbers to protect against replay attacks.
\ac{ESP} \cite{rfc4303} protects the confidentiality of \ac{IP} packets by symmetric encryption.
As for \ac{AH}, it also adds sender authentication, packet integrity validation, and protection against replay attacks.
\ac{ESP} supports symmetric ciphers such as \ac{3DES}, Blowfish, and \ac{AES}.
Ciphers that only apply encryption are combined with an authentication function.
\ac{AES} in \ac{CBC} or \ac{CTR} mode are examples for such ciphers that might be combined with \ac{SHA} for authentication.
\ac{AE} ciphers such as \ac{AES} in \ac{GCM} \cite{rfc4106} include both, packet encryption and authentication.
IPsec provides support for \ac{IPComp} \cite{rfc3173} so that the payload of \ac{IP} packets can be compressed before encryption.

\subsubsection{Operation Modes}
\begin{figure}[ht]
    \begin{center}
    \includegraphics[width=0.99\linewidth]{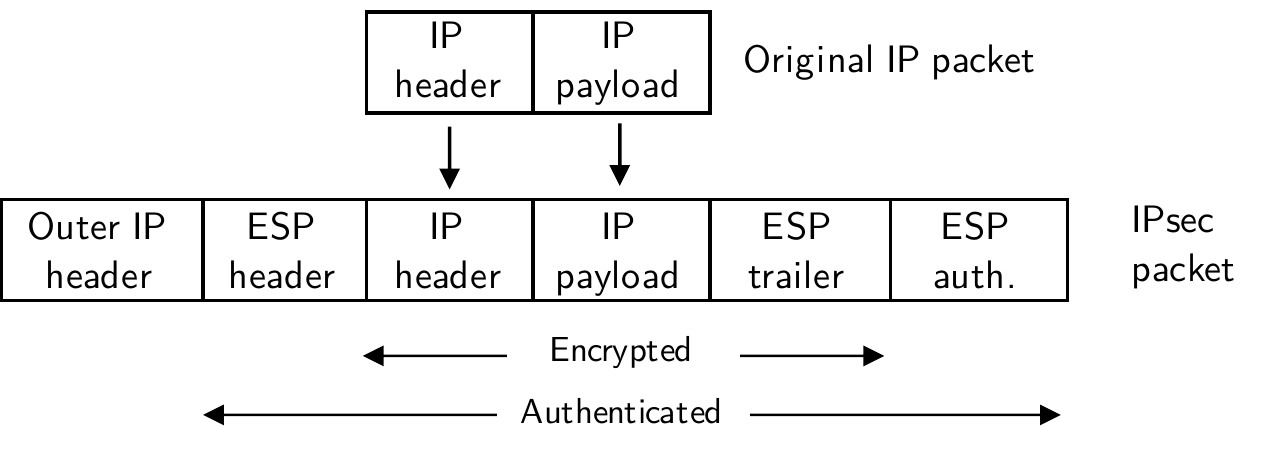}
    \end{center}
    \caption{Tunnel mode with \ac{ESP}. The original \ac{IP} packet is inserted between the \ac{ESP} header and the \ac{ESP} trailer. The inner \ac{IP} packet is encrypted while the complete \ac{ESP} packet is authenticated.}
    \label{fig:ipsec-tunnel}
\end{figure}

\ac{IPsec} can be deployed in either \textit{transport} or \textit{tunnel} operation mode.
Transport mode protects \ac{IP} traffic that is exchanged between two network hosts (host-to-host scenario).
An \ac{AH} or \ac{ESP} header is inserted between the \ac{IP} header and the \ac{IP} payload.
Tunnel mode protects \ac{IP} traffic host-to-host, host-to-site, and site-to-site communication scenarios.
\fig{ipsec-tunnel} depicts how tunnel mode with \ac{ESP} is applied to an \ac{IP} packet.
A new outer \ac{IP} header with the \ac{IP} addresses of the \ac{IPsec} peers is created.
The original \ac{IP} packet is inserted between the \ac{ESP} header and the \ac{ESP} trailer.
Encryption protects the original \ac{IP} packet while authentication is applied to the complete \ac{ESP} packet.

\subsubsection{Core Components}
We describe the core components of \ac{IPsec} implementations that are part of hosts or gateways.
The \textit{\acf{SPD}} holds security policies that decide on traffic protection using \ac{IPsec}.
Entries have match keys, e.g., \ac{IP} src/dst address, \ac{IP} protocol, and TCP/UDP port, with an assigned action.
\ac{IPsec} allows three actions: DROP (discard packet), BYPASS (no protection), and PROTECT (apply \ac{IPsec} protection).
In case the table yields no match, the DROP action is applied.
\ac{SPD} entries for \ac{IPsec} connections point to the protocol (\ac{AH}/\ac{ESP}), the operation mode (transport/tunnel), and the cipher suite.
An \ac{IPsec} tunnel between two peers is described by two unidirectional \acfp{SA}.
An \ac{IPsec} \ac{SA} contains all required data for \ac{AH}/\ac{ESP} processing, e.g., cipher keys, valid sequence numbers, or \ac{SA} lifetimes for rekeying and tear down.
\acp{SA} are part of the \textit{\acf{SAD}}.
With the information from the \ac{SAD}, packets then can be processed by \textit{\ac{ESP}/\ac{AH} processing}.
Although manual configuration of \ac{SA} is possible, \acp{SA} are typically configured between \ac{IPsec} peers with the help of the \acf{IKE} protocol \cite{rfc2409} that was introduced with \ac{IPsec}.
It authenticates both peers, sets up a secure channel for key exchange, and negotiates \acp{SA}.
Today, its successor \ac{IKEv2} \cite{rfc7296} should be used.
It is less complex and solves incompatibility issues of \ac{IKE}.
\ac{IKE} relies on the \textit{\acf{PAD}} for authentication other \ac{IPsec} peers.

\begin{figure}[ht]
    \begin{center}
    \includegraphics[width=.85\linewidth]{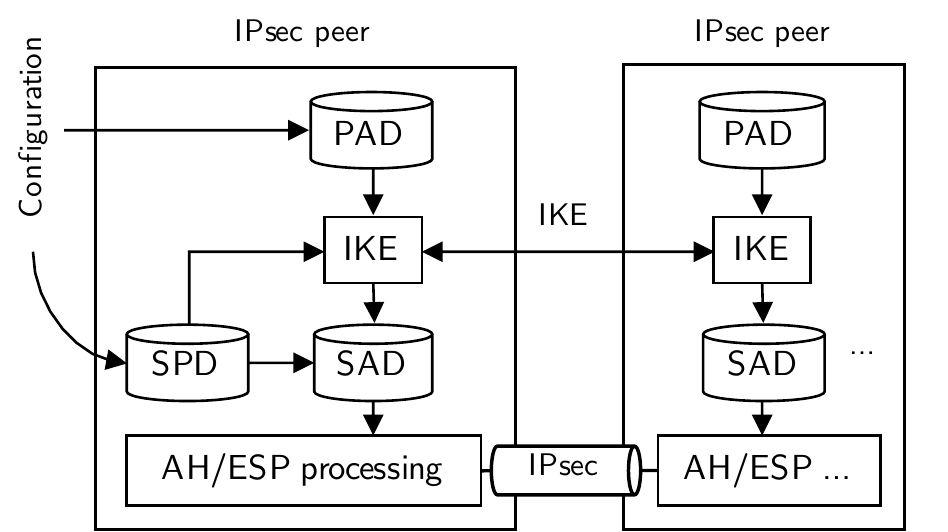}
    \end{center}
    \caption{\ac{IPsec} packet processing between two \ac{IPsec} peers. Each peer features a \ac{SPD}, \ac{SAD}, \ac{PAD}, and \ac{AH}/\ac{ESP} processing functions on the data plane. The \ac{SPD} and \ac{PAD} are configured manually where \ac{SAD} entries are managed by the \ac{IKE} daemon.}
    \label{fig:ipsec-components-interaction}
\end{figure}

\subsubsection{Packet Processing}
\begin{figure}[ht]
    \begin{center}
    \includegraphics[width=.99\linewidth]{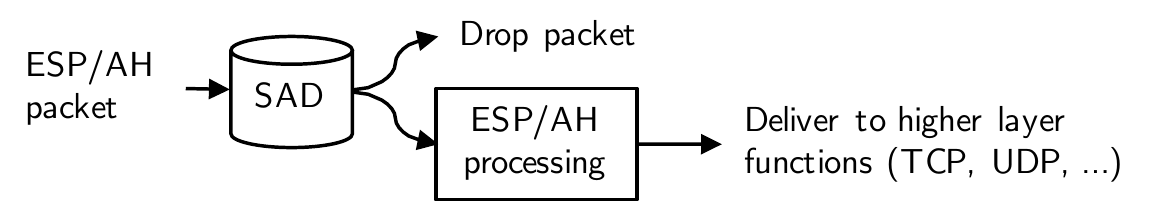}
    \end{center}
    \caption{\ac{IPsec} ingress processing. Arriving packets with an \ac{ESP}/\ac{AH} header are processed with the help of the \ac{SAD}. \ac{ESP}/\ac{AH} processing relies on data in the \ac{SAD}. In case of no match, the packet is dropped.}
    \label{fig:ingress-processing}
\end{figure}

\ac{IPsec} differentiates between ingress and egress processing of packets.
\fig{ingress-processing} depicts ingress processing.
Arriving packets that have an \ac{ESP}/\ac{AH} header are processed with the help of the \ac{SAD}.
If the \ac{SAD} has an entry for the corresponding \ac{SA}, the \ac{SA} data is forwarded to the \ac{ESP}/\ac{AH} processing function that removes \ac{IPsec} protection.
Afterwards, the packet is forwarded to default network processing.

\begin{figure}[ht]
    \begin{center}
    \includegraphics[width=.99\linewidth]{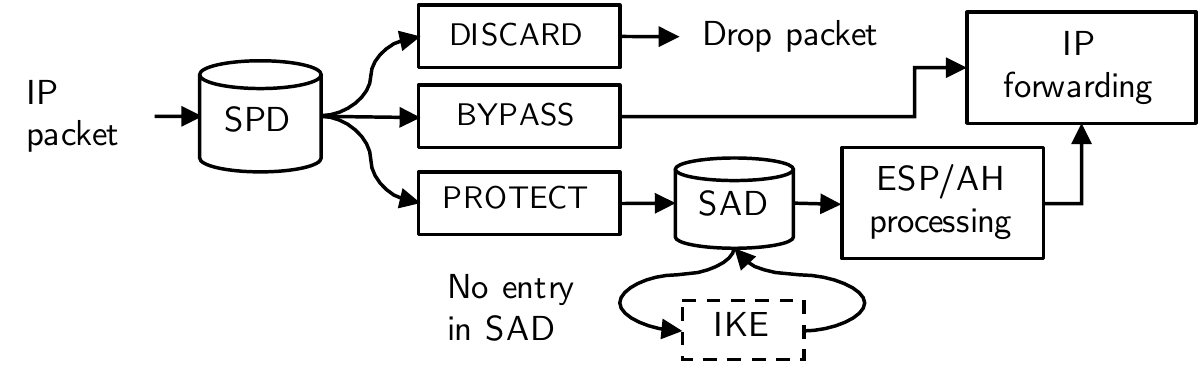}
    \end{center}
    \caption{\ac{IPsec} egress processing. The \ac{SAD} matches packets and maps them to the actions DISCARD, BYPASS, and PROTECT. In case of BYPASS, the packet is passed to IP forwarding. In case of PROTECT, the \ac{SAD} is searched for a corresponding entry for \ac{ESP}/\ac{AH} processing. In case of no match, the packet is dropped.}
    \label{fig:egress-processing}
\end{figure}

\fig{egress-processing} depicts egress processing where \ac{IP} packets are matched with \ac{SPD} entries as explained before.
In case of PROTECT, data for \ac{ESP}/\ac{AH} processing is selected from the \ac{SAD}.
If the \ac{SAD} has no matching entry, \ac{SA} setup is requested from the \ac{IKE} daemon.

\subsubsection{Discussion}
Among more recent alternatives such as OpenVPN and WireGuard, \ac{IPsec} is still one of the most widespread \ac{VPN} mechanisms nowadays.
\ac{IPsec} implementations are part of common operating systems for computers, servers, and mobile devices for many years.
Most network hardware appliances, e.g., firewalls, routers, or security appliances, include an \ac{IPsec} implementation.

However, \ac{IPsec} is highly criticized for its complexity for many years.
The most encompassing analysis was performed by Ferguson and Schneier \cite{schneier-ipsec} in 2003.
The authors mainly criticized the redundancy of functionality caused by \ac{AH}, \ac{ESP}, and the two operation modes, the complex key exchange with \ac{IKE}, and the complex configuration caused by the \ac{SPD} and \ac{SAD}.
However, those issues can be easily solved.
Transport mode and \ac{AH} should be avoided.
Instead, \ac{AE} ciphers that combine encryption and authentication should be used in conjunction with \ac{ESP} with tunnel mode.
\ac{IKE} should be substituted by a less complex protocol for key exchange.
In P4-IPsec, we follow those recommendations and restrict the \ac{IPsec} implementation to \ac{ESP} in tunnel mode with controller-based \ac{SA} management without \ac{IKE}.

\subsection{Data Plane Programming with P4}
\label{sec:p4-foundations}
\ac{SDN} introduces network programability by shifting control-plane functions to a software-based controller that determines the packet processing behaviour of network devices.
\acf{OF} \cite{openflow-paper} is the most widely-used \ac{SDN} approach.
It relies on data plane devices with a fixed set of functions and a southbound interface to the \ac{SDN} controller.
The \ac{SDN} controller defines how these functions are applied to network packets.
Programmable data planes extend network programmability to data plane functionality.
Packet processing can be defined on an abstract layer using a dedicated programming language.
Thereby, packet processing behavior is decoupled from the underlying hardware.
This new principle facilitates open network research with support for agile development processes and flexible deployment options.
Bifulco \etal{}\cite{BiRe18} give an overview on programmable data planes.
Target platforms include software targets, \acp{NIC}, \acp{NIC} with a \ac{FPGA} unit, and hardware appliances with \acp{NPU}.
P4 is the most widely-used data plane programming language nowadays.
Initially presented as a research paper in 2014, the project is now standardized by the P4 Language Consortium under the \ac{ONF}.
Its latest specification is version 16 (P4\textsubscript{16}) \cite{p4-specification}.

\subsubsection{Processing Pipeline}
\begin{figure}[ht]
    \begin{center}
    \includegraphics[width=.99\linewidth]{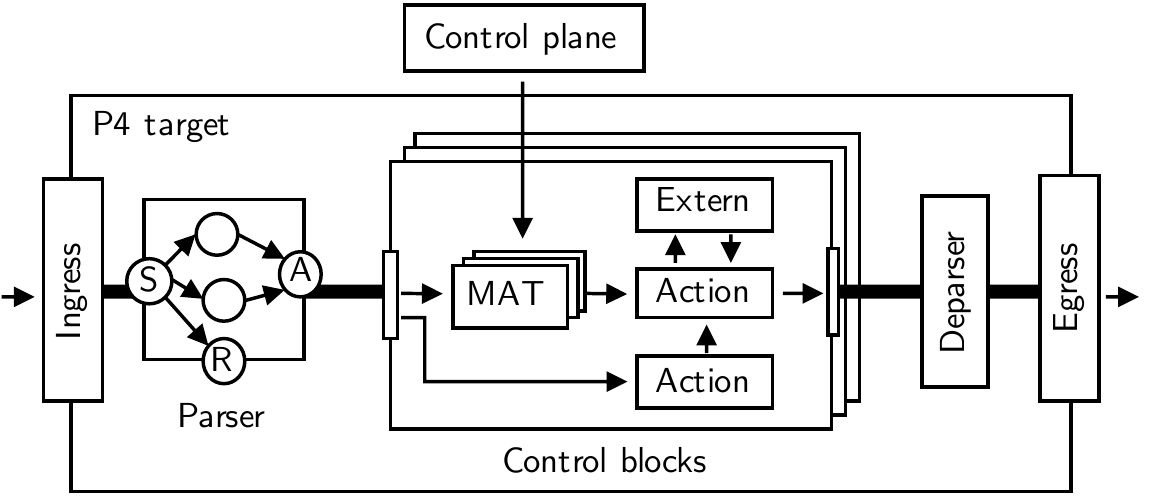}
    \end{center}
    \caption{Simplified view on the P4 processing pipeline of P4\textsubscript{16}. It comprises the parser, control blocks, and the deparser. Each control block may include \acp{MAT}, actions, and externs.}
    \label{fig:p4-processing-pipeline}
\end{figure}

\fig{p4-processing-pipeline} depicts a simplified view on the packet processing pipeline of P4.
It consists of three core abstractions that help to express forwarding behavior.

\paragraph{Parser}
The parser extracts header fields of packets into internal data structures.
P4 does not include predefined header types, i.e., programmers need to define packet formats and extraction behavior.
Packet header formats are defined using P4 header types such as fixed- and variable-length bit strings or integers.
The extraction behavior of the parser is expressed as \ac{FSM}.
Parsing is initiated in the state \textit{start}, possible outcome states are \textit{accept} (proceed in packet processing) and \textit{reject} (drop the packet).
Custom states that are positioned between start and ending states implement the extraction of header data.
Transitions between those states are formulated using conditions.
For example, after successfully parsing an \ac{IP} header, state transitions to TCP or UDP parsing might follow.

\paragraph{Control Blocks}
Control blocks are functions that modify packet headers and metadata.
The P4 processing pipeline can include multiple control blocks that are typically separated by a queue or buffer.
Packet processing in control blocks is stateless: the outcome of packet processing applied to one packet can not influence packet processing applied on a subsequent packet.
Actual packet processing is implemented in \textit{actions}, code fragments within control blocks that implement read/write operations with functions provided by P4, e.g., setting header fields or adding/removing headers.
Actions can be called from other actions, explicitly with the start of the control block, or implicitly by \textit{\acp{MAT}}.
\acp{MAT} map match keys to particular actions with associated parameters.
When applying a \ac{MAT} to packets, header and metadata is matched in exact, ternary, or in longest prefix manner against the keys of the \ac{MAT}.
If matching yields a particular row entry, the specified action is called with the associated parameters.
If there is no match, a default action is applied.
P4 programs only contain the declaration of \acp{MAT}, their entries are maintained by a control plane via an \ac{API} in runtime.
Some targets may provide additional functions for packet processing, e.g., particular functions such as checksum generation, or stateful components such as counters, meters, and registers.
These components can be used within P4 programs as so-called \textit{externs}.
Externs have an interface with defined instantiation methods, functions, and parameters.
After import and declaration, they can be used in control blocks just like any other P4 function.

\paragraph{Deparser}
The deparser reassembles the packet header and payload and serializes it to be sent out via an egress port.

\subsubsection{Deployment Model}
Software or hardware platforms that execute P4 programs are called P4 targets.
Common software targets are the BMv2 \cite{bmv2} software target, eBPF packet filters, and the T\textsubscript{4}P\textsubscript{4}S \cite{t4p4s} software target that includes hardware interfaces via \ac{DPDK} \cite{dpdk} and \ac{ODP} \cite{odp}.
Hardware targets include \ac{FPGA}-based targets and \acp{NIC}, \ac{NPU}-based \acp{NIC}, and whitebox switches featuring the Tofino \ac{ASIC} from Barefoot Networks.
P4 programs are implemented for a particular P4 architecture.
P4 architectures can be seen as programming models that represent the logical view of a P4 processing pipeline.
They serve as intermediate layer to decouple P4 programs from P4 targets, i.e., P4 programs that are implemented for a particular P4 architecture can be deployed to all P4 targets that implement this architecture.
A front-end compiler translates P4 programs into a target-independent \ac{HLIR}.
Afterwards, the \ac{HLIR} is compiled to the particular target using a back-end compiler that is provided by the manufacturer.

\subsubsection{Control Plane API: P4Runtime}
\label{foundations:p4-runtime}
The runtime behavior of P4 targets can be controlled by managing \acp{MAT} or stateful components (e.g., counter, meters, registers, or externs) that are part of the P4 program.
P4Runtime \ac{API} \cite{p4-runtime-specification} is a target- and program-independent \ac{API} standardized by the P4 language consortium.
P4Runtime uses gRPC for communication between the control plane and P4 targets and protobuf \cite{protobuf} data structures for packet serialization/parsing.
gRPC connections can be secured with \ac{TLS} and mutual authentication with certificates.
In P4Runtime, the \ac{SDN} controller establishes gRPC connections to pre-configured targets.
P4Runtime supports P4 object access (e.g., on \acp{MAT} and externs), session management (master/slave controllers), role-based access control, and a packet-in/-out mechanism to receive and send out packets via controllers.
The PI Library is the reference implementation of the P4Runtime server that is part of P4 targets.
It implements generic functionality for internal P4 objects such as \acp{MAT}.
This functionality can be extended by target- or architecture-specific configuration objects.
\emph{p4runtime\_lib} \cite{p4-runtime-lib} is an exemplary implementation of the P4Runtime \ac{API} in Python to be used for building controllers.
P4Runtime \ac{API} plugins are also available for common \ac{SDN} controllers such as ONOS or OpenDaylight.

\subsubsection{Application Domains}
Most research works on P4-based network applications target data center or wide area networks.
In traffic management and congestion control, P4 is leveraged to implement new congestion notification mechanisms, novel traffic scheduling mechanisms, or novel mechanisms for active queue management.
In routing and forwarding, special routing and forwarding mechanisms, publish-subscribe systems, or novel concepts from the area of named data networks are implemented.
A large focus also lies on monitoring, where several works implement monitoring systems, sketch-based monitoring mechanisms, and \ac{INT} systems.
Besides, P4 is used in data center scenarios to implement switching, load balancing, \ac{NFV}, and \ac{SFC} mechanisms.
\section{Related Work}
\label{sec:related-work}

We describe related work on network security applications built with P4, \ac{IPsec} in \ac{SDN}, and implementation of \ac{IPsec} packet processing.

\subsection{Network Security Applications with P4}
Although network security is not the prevalent application domain of P4, some scientific work has been published in this field.
We describe related work on firewalls, DDoS mitigation mechanisms, and other security applications.

\subsubsection{Firewalls}
Vörös and Kiss \cite{VoKi16} introduce a P4-based firewall for filtering IPv4, IPv6, TCP, and UDP packets.
It includes a ban list for instant drop, counters, e.g., for measuring the packet rate or unsuccessful connection attempts, and \acp{MAT} for applying whitelist firewall rules.
P4Guard \cite{DaCh18} follows a similar approach.
Its authors focus on simplified updated processes by deploying re-compiled versions of the P4 program.
Ricart-Sanchez \etal{}\cite{RiMa18} implement a P4-based firewall for 5G networks.
It includes parser definitions for filtering GPRS tunneling protocol (GTP) data.
CoFilter \cite{CaBi18} introduces a hash function for efficient flow identification.
It is built as P4 action and uses hashes instead of 5-tuples for flow identification to save table space.
Including the function directly on the packet processing devices keeps latency low.
Zaballa \etal{} \cite{ZaFr20} and Almaini \etal{} \cite{AlAl19} introduce port knocking on P4 switches.

\subsubsection{DDoS Mitigation Mechanisms}
Paolucci \etal{}\cite{PaCu18, PaCi19} propose a DDoS mitigation mechanism that runs on P4 switches.
A stateful mechanism detects and blocks DDoS port scan attacks with incremental TCP and UDP destination port numbers.
Dimolianis \etal{} \cite{DiPa20} also implement a DDoS attack mitigation mechanism that runs completely on P4 switches.
Collected flow data is mapped to distinct time intervals where DDoS attacks are detected by analyzing the symmetry ratio of incoming and outgoing traffic.
TDoSD@DP \cite{FeXi18} implements a mitigation scheme against DDoS attacks on SIP proxies.
The authors introduce a simple state machine that monitors SIP message sequences.
Valid sequences of INVITE and BYE messages keep the port open.
Febro \etal{} \cite{FeXi19} implement another DDoS mitigation mechanism for SIP INVITE DDoS attacks.
P4 switches keep per-port counters for INVITE or REGISTER packets that are monitored by an SDN controller to detect DDoS attacks.
LAMP \cite{GrLi18} implements cooperative mitigation of application layer DDoS attacks via in-band signaling with P4.
Afek \etal{} \cite{AfBr17} implement known mitigation mechanisms for SYN and DNS spoofing in DDoS attacks in P4.
Lapolli \etal{} \cite{LaMa19} describe a novel algorithmic approach based on the Shannon entropy to detect and stop DDoS attacks on P4 switches.
Kuka \etal{}\cite{KuVo19} introduce an FPGA-based system for DDoS attack mitigation.
P4 is used to extract header data from packets and send it to an SDN controller where DDoS attack identification is implemented.
Mi and Wang \cite{MiWa19} propose a similar approach where collected data is sent to a deep learning module that runs on a server in the network.

\subsubsection{Other Security Applications}
Lewis \etal{} \cite{LeBr19} implement an IDS offloading mechanism in P4.
A rule parser translates Snort IDS rules into MAT entries for a P4 switch.
Then, IDS pipeline stages decide if packets should be forwarded, dropped, or sent to an external IDS for analysis.
Poise \cite{KaXu20} is a security-related network control system that translates high-level policies into P4 programs for network control.
In P4-MACsec \cite{p4-macsec}, we implement IEEE 802.1AE (MACsec) in P4 and introduce an automated deployment that relies on link monitoring and MACsec provisioning.
Link monitoring is implemented using a novel variant of \ac{LLDP} that relies on encrypted payloads and sequence numbers to protect against \ac{LLDP} packet manipulations and replay attacks.

\subsection{IPsec in SDN}
Several works investigate the application of \ac{SDN} to \ac{IPsec} operation.
We describe operation modes, southbound interfaces, and use cases.

\subsubsection{Operation Modes}
\label{sec:operation-modes}
\begin{figure}[ht]
    \begin{center}
    \includegraphics[width=.92\linewidth]{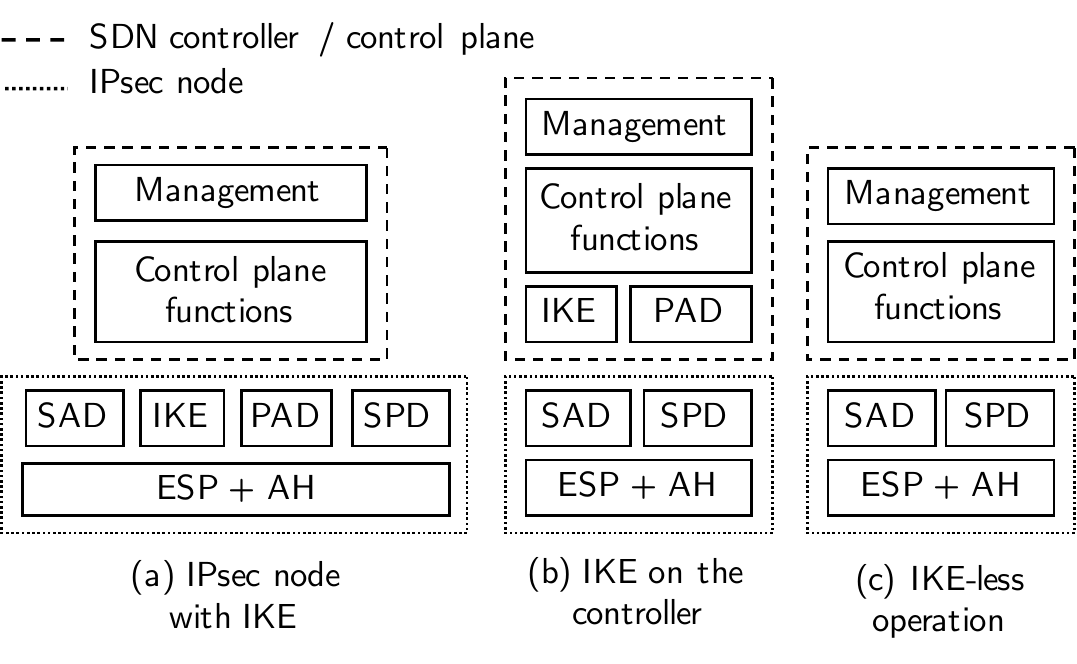}
    \end{center}
    \caption{Operation Modes for Data Plane Management of \ac{IPsec}. In (a), \ac{IKE} is part of the \ac{IPsec} node. In (b), \ac{IKE} is part of the control plane. In (c), \ac{IKE} is substituted by controller-based \ac{SA} management.}
    \label{fig:ike-scenarios}
\end{figure}

Related work can be categorized by three different operation modes that are depicted in \fig{ike-scenarios}.

\paragraph{IPsec Node with IKE}
In the first operation mode, \ac{IPsec} processing nodes feature an \ac{IKE} daemon, \ac{SDN} assists in preconfiguration.
Aragon \etal \cite{ArTi18, aragon-ace-ipsec-profile-01} propose that an \ac{SDN} controller pre-configures authentication keys in the \ac{PAD}.
Carrel and Weiss \cite{draft-carrel-ipsecme-controller-ike-01} propose that an \ac{SDN} controller distributes Diffie-Hellman public values to all associated \ac{IPsec} data plane nodes.
Guo \etal \cite{GuYa03} propose a similar approach that is compatible to older \ac{IKE} daemons that only support IKEv1.
Lopez-Millan \etal \cite{LoMa19} propose an "\ac{IKE} mode" where the \ac{SDN} controller only provides information for configuration of \ac{SPD}, \ac{PAD}, and \ac{IKE} daemon.
All proposals aim to reduce the message exchanges in an \ac{IKE} process by preconfiguring it by a controller.

\paragraph{\ac{IKE} on the Controller}
In the second operation mode, the \ac{IKE} daemon is part of a \ac{SDN} control plan.
Son \etal \cite{SoXi17} relocate the \ac{IKE} daemon to the control plane.
There, it performs key exchange with peers and manages the \ac{SAD} of the \ac{IPsec} data plane nodes.
This approach even supports migration schemes so that the \ac{SA} can be transferred to other \ac{IPsec} data plane nodes, e.g., in fail-over or load-balancing operations.
Vajaranta \etal \cite{VaKa17} describe a similar approach where \ac{IKE} is executed as network function that can be scaled up by creating additional instances.

\paragraph{\ac{IKE}-less Operation}
In the third operation mode, \acp{SA} are maintained without \ac{IKE}.
Lopez-Millan \etal \cite{LoMa19} describe an "\ac{IKE}-less" operation mode where the \ac{SA} maintenance is delegated to an \ac{SDN} controller.
Here, the \ac{IPsec} element only implements \ac{IPsec} logic where the complete key management logic is moved to the \ac{SDN} controller.
As there is no \ac{IKE}, no \ac{PAD} is required.
The authors differentiate between a proactive mode, where \ac{SPD} and \ac{SAD} are preconfigured by the \ac{SDN} controller and reactive mode, where only the \ac{SPD} is preconfigured by the \ac{SDN} controller.
Several works \cite{ArTi18,ietf-i2nsf-sdn-ipsec-flow-protection-07,Br17,sajassi-bess-secure-evpn-02} propose \ac{SA} management without \ac{IKE}.
The controller generates keying material and sets up \acp{SA} in the \ac{SAD} of associated \ac{IPsec} data plane nodes.
Gunleifsen \etal \cite{GuGk18} introduce a key management server that creates and distributes \ac{IPsec} \acp{SA} for encryption \acp{VNF}.
In consecutive works, Gunleifsen \etal \cite{GuKe19a, GuKe19b} name this concept as "Software-Defined Security Associations (SD-SAs)".
Encryption \acp{VNF} only perform \ac{IPsec} processing.
\acp{SA} are created and distributed by an authentication center.

\subsubsection{Discussion on Operation Modes}
We briefly discuss benefits and drawbacks of the three operation modes.
The first operation mode benefits from easy migration.
As legacy \ac{IPsec} devices already feature an \ac{IKE} daemon, they can be easily extended by an interface to profit from \ac{SDN}-assisted operation of \ac{IPsec} (see \cite{LoMa19}).
The second operation mode especially introduces flexibility and scalability.
Separating \ac{IPsec} processing and \ac{SA} establishing to different entities improves scalability (see \cite{VaKa17}).
The third operation mode removes the overhead of peer-to-peer key exchange with \ac{IKE}.
On the one hand, this might be unnecessary in environments where both \ac{IPsec} peers are controlled by an \ac{SDN} controller.
On the other hand, \ac{IKE} requires that \ac{IKE} connectivity of both peers which could be not given in particular scenarios (see \cite{GuKe19b}).
Lopez-Millan \etal\cite{LoMa19} show in an analytical evaluation that \ac{IKE}-based and \ac{IKE}-less operation of IPsec have the approximately same process load in terms of messages and configuration data exchange.

\subsubsection{Southbound Protocols}
On legacy network devices that feature \ac{IPsec} devices, SNMP (e.g., \cite{cisco-ipsec-mib}) is used for basic configuration and monitoring.
The authors of \cite{GuYa03} extend this usage in making an \ac{IKE} daemon manageable by SNMP as well.
In \cite{AlBa18}, SSH is used as southbound interface to manage and monitor \ac{IPsec} data plane nodes.
The work in \cite{ietf-i2nsf-sdn-ipsec-flow-protection-07} uses NETCONF with YANG configuration models.
In addition to the southbound protocol, they consider east-/westbound interfaces for controller-to-controller communication via different domains.
Aragon \etal{}\cite{ArTi18} used OAuth 2.0 to deliver configuration data within authorization messages.
In \cite{Br17}, \acf{OF} is extended using experimenter messages.
The work in \cite{sajassi-bess-secure-evpn-02} leverages BGP.
Li and Mao \cite{LiMa15} use a custom southbound protocol to interface an \ac{IPsec} extension module on an Open vSwitch.
The authors of \cite{LiLi16} propose a custom southbound protocol with notification, configuration, and query messages that are transmitted via TCP or TLS.
Lopez-Millan \etal \cite{LoMa19} use NETCONF with YANG models as southbound protocol.
Gunleifsen \etal \cite{GuKe19a, GuKe19b} use REST with JSON.

\subsubsection{Use Cases}
Use cases that benefit from controller-based operation of \ac{IPsec} are SD-WAN, cloud provider networks, and dynamic \ac{VPN} setup.

\paragraph{SD-WAN}
Large organizations with distributed locations require network connectivity between the different sites.
As dedicated links are expensive, site-to-site IPsec-VPNs over provider networks are increasingly used.
However, manually setting up \ac{VPN} connections between all branches is time-intensive and complex.
SD-WAN \cite{AlBa18, GuYa03, LiMa15} proposes \ac{IPsec} data plane functionality as part of hardware appliances or software modules at the perimeter of the different sites of the organization.
Then, a centralized controller automatically sets up and maintains IPsec-VPN connections.

\paragraph{Cloud Provider Networks}
Often, internal services offered by a public or private cloud provider need to be accessed from within networks of an organization.
Again, site-to-site IPsec-VPN tunnels are a cost-efficient alternative to dedicated links.
Administrators define IPsec-VPN gateways via a cloud management interface.
Then, the cloud orchestrator deploys IPsec-VPN gateways as virtual network function on the cloud provider's infrastructure.
Its runtime operation is managed by a controller.
In addition, controller-based operation of \ac{IPsec} can be also used to dynamically connect different cloud networks by a multi-cloud orchestrator \cite{MeHo13}.
Gunleifsen \etal \cite{GuGk18, GuKe19a} propose hop-by-hop protection for \acp{SFC} using IPsec and controller-based operation.

\paragraph{Dynamic VPN Setup}
Managing many IPsec-VPN connections to different hosts or services on a client host can be cumbersome.
Dynamic \ac{VPN} setup performed by a controller takes over the tasks of tunnel setup and management.
Van der Pol \etal \cite{PoGi16} present a concept where users request \ac{VPN} access to a particular network device from the controller.
It then automatically sets up a \ac{VPN} tunnel to the remote domain.
Aragon \etal \cite{ArTi18} combine dynamic \ac{VPN} setup with authentication and authorization to automatically deploy IPsec-VPN tunnels between IoT network devices.
This introduces several advantages over traditional deployment.
First, the control plane has an encompassing view on the network topology with all devices.
It can monitor usage and detect outages for reliable operation.
Second, the centralized control plane features northbound interfaces for management applications and southbound interfaces for controlling data plane devices.
Instead of manual per-device configuration, \acp{VPN} are operated via a management layer with policy languages that allow rule validation.
Last, the centralized control plane offers flexibility so that \ac{VPN} operation can be extended by other mechanisms, e.g., user authentication with 802.1X \cite{LiMa15}.

\subsection{Implementation of IPsec Packet Processing}
\label{sec:ipsec-packet-processing}

With P4-IPsec, we present the first data plane implementation of \ac{IPsec} in P4.
We give an overview on \ac{IPsec} data plane implementations as related work.

\subsubsection{Software Implementations with Hardware Acceleration}
\ac{IPsec} software programs represent the most simple packet processing implementations.
Their I/O performance depends on the hardware, the chosen cryptographic algorithms, and the average packet size.
For Linux host systems, optimization techniques such as DPDK \cite{dpdk}, Netmap \cite{Ri12}, and PF\_RING \cite{pf-ring} tweak network stack processing to increase packet I/O rates.
Other works propose to increase \ac{IPsec} packet I/O by using multiple CPU cores \cite{LiHu18,XiHa13} or the GPU \cite{HaJa10}.
Gallenmüller \etal{}\cite{GaEm15} compare several mechanisms in an extensive study.
Most of the described optimization mechanisms are only applicable to Linux operating systems.

\ac{IPsec} packet I/O of software implementations can be improved by offloading crypto operations or \ac{IPsec} operations to hardware.
For the former, current CPU architectures provide hardware acceleration for common crypto operations.
AES-NI \cite{aes-ni} or ARMv8 Cryptographic Extensions \cite{armv8-architecture} are examples of \ac{AES} instruction sets that replace pure software implementations.
System on chip \acused{SoC}(\ac{SoC}) platforms or circuit boards may contain chips for offloading cryptographic processing.
Examples are the Marvell Cryptographic Engines Security Accelerator (CESA) or Intel QuickAssist \cite{DiRi13}.
Such processors can be also part of extension circuit boards that are connected to the mainboard via PCI.
\acp{FPGA} might be also used for implementing crypto operations, several vendors (e.g., \cite{algotronix-aes-ip-core}) supply implementations of cryptographic algorithms as program cores.
For the latter, IPsec hardware accellerators are available as \ac{ASIC} \cite{HaLe04,HoSc04}, \ac{NPU}\cite{LiXu08, MeCh11}, \ac{APU} \cite{PaJu16}, or \ac{FPGA} \cite{RaCo16, VaVi18}.

\subsubsection{Hardware Implementations}
Proprietary \ac{IPsec} hardware concentrators, e.g., as sold by Cisco or Juniper, are optimized for high \ac{IPsec} I/O rates and, therefore, might implement a larger degree of the overall \ac{IPsec} processing operations in hardware (e.g., \acp{ASIC}).
Due to their disclosed architectural details, we cannot get insight into technical details.
In addition, encompassing \ac{IPsec} implementations for \acp{FPGA} exist \cite{KoSk17,DrGu12} where only \ac{SPD} and \ac{SAD} are managed by an \ac{SDN} controller.

\subsubsection{Implementations on Programmable Data Planes}
For programmable data planes, in 2016, a Xilinx employee reported on the P4-Development mailing list \cite{p4-mailinglist-ipsec} that IPsec was successfully implemented in PX \cite{sdnet-px}, a high-level domain-specific programming language for programmable data planes.
Crypto primitives are not expressed in the language, but by an extern mechanism similar to P4's externs.
The authors report that the crypto primitives were programmed as \ac{RTL} designs targeting \acp{FPGA}.
The authors report that the principle should be exactly the same for P4, but it was not ported so far.
\section{Concept}
\label{sec:concept}

We describe the concept of P4-IPsec.
We give an overview, discuss design choices, and describe its data plane and control plane in detail.

\subsection{Overview}

\begin{figure}[ht]
    \begin{center}
    \includegraphics[width=.99\linewidth]{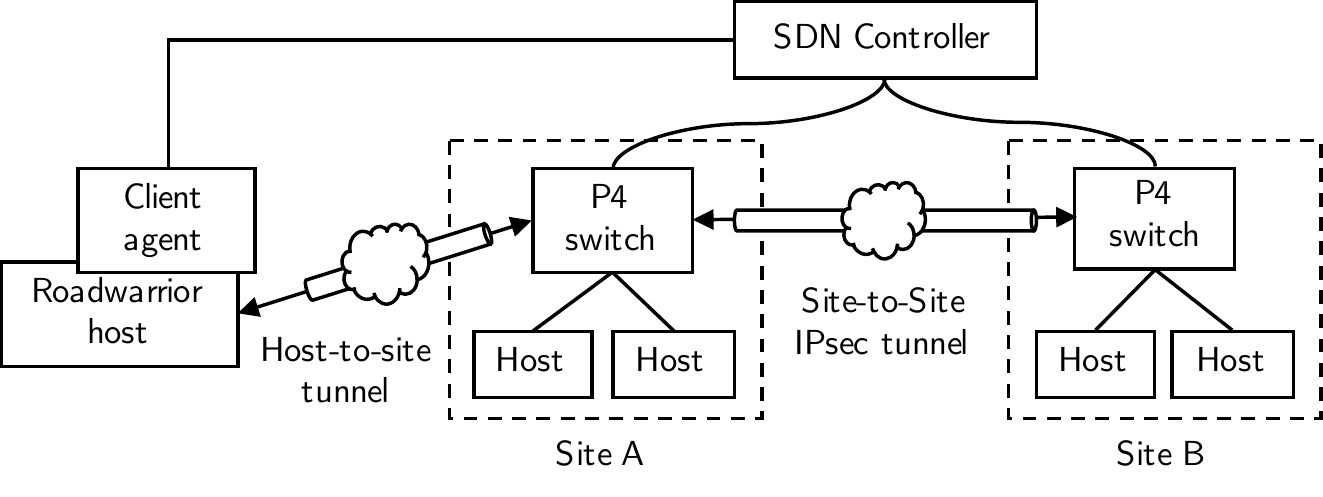}
    \end{center}
    \caption{Overview on the functionality of P4-IPsec. In host-to-site operation, roadwarrior hosts run a client agent to setup an \ac{IPsec} tunnel to a P4 switch via the \ac{SDN} controller. In site-to-site operation, the \ac{SDN} controller sets up \ac{IPsec} tunnels on pairs of P4 switches.}
    \label{fig:overview}
\end{figure}

\fig{overview} gives an overview on the functionality of P4-IPsec.

\begin{figure*}[ht]
    \begin{center}
    \includegraphics[width=\linewidth]{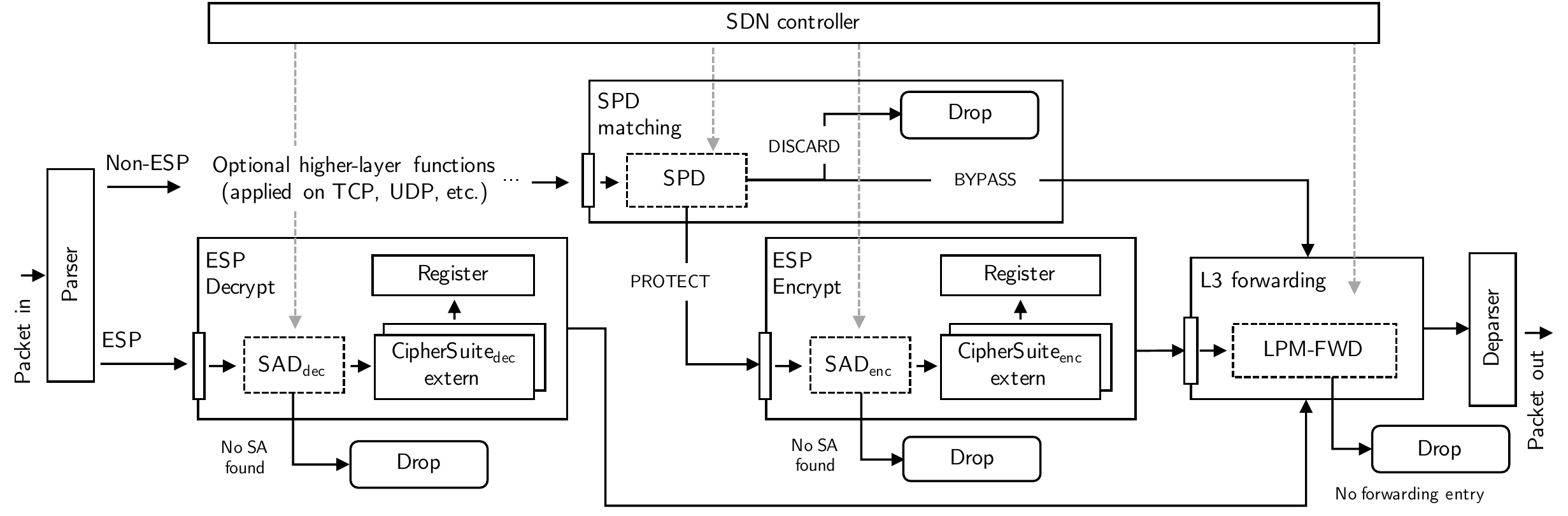}
    \end{center}
    \caption{Data plane processing pipeline of P4-IPsec. For ease of understanding, related functionalities are grouped together as function block.}
    \label{fig:data-plane-overview}
\end{figure*}

P4-IPsec supports two IPsec tunnel operation modes: \textit{host-to-site} and \textit{site-to-site}.
In host-to-site mode, roadwarrior hosts establish IPsec tunnels to access internal networks.
Roadwarrior hosts run a client agent that interacts with the controller for tunnel setup.
In site-to-site mode, two internal networks are connected via an IPsec tunnel that is established between two P4 switches.
As core principle of P4-IPsec, every P4 switch implements the same IPsec functionalities, i.e., it can act as both, IPsec tunnel endpoint for roadwarrior hosts in host-to-site mode and for other P4 switches in site-to-site mode.
This facilitates very flexible deployments where IPsec tunnels do not necessarily terminate at a central VPN concentrator but can be distributed to many P4 switches instead.

\subsection{Design Choices}
P4 programs describe the packet forwarding behavior of switches or routers.
Thereby, an implementation of IPsec in P4 is limited to data plane centric parts.
Additional mechanisms such as \ac{IKE} need to be part of an \ac{SDN} controller implementing the control plane and interfacing the P4 program.
For P4-IPsec, our adoption of IPsec in P4, we make the following design choices:

\subsubsection{Use of IKE-less Operation Mode}
Refering to the results of Lopez-Millan et al. \cite{LoMa19} (see \sect{operation-modes}), we choose to implement \ac{SA} management via the \ac{SDN} controller without \ac{IKE}.
Our proposed P4 processing pipeline comprise equivalent representations for the \ac{SAD} and \ac{SPD} that are both maintained by the SDN controller.
Due to the lack of \ac{IKE}, no \ac{PAD} is required on the P4 processing pipeline.
Selecting IKE-less operation mode does not exclude an integration of \ac{IKE} on the SDN controller at a later stage.

\subsubsection{Restriction to ESP in Tunnel Mode}
To keep our proposed concept as minimalistic as possible, we adopt the recommendations of Ferguson and Schneier \cite{schneier-ipsec} and restrict our implementation to \ac{ESP} in tunnel mode.

\subsubsection{Implementation of Cipher Suites with Externs}
P4 does not provide functions for encryption, decryption, and message authentication.
In contrast to related work, we decide against offloading IPsec processing to external, software-based processing nodes and implement cipher suites with the help of P4 externs (see \sect{p4-foundations}).
This should decrease the latency introduced by external processing while keeping the overall system more minimalistic.
Each cipher suite is implemented by two externs; one that implements encryption functionality, and one that implements decryption functionality.

\subsubsection{Prototype Simplifications}
We limit P4-IPsec to IPv4 and omit support for IPv6.
We also omit support for \ac{IPComp}.
For applicability in experiments, we implement simple L3 forwarding based on \ac{LPM}.
Clearly, this is not a requirement from the IPsec standard.

\subsection{Data Plane of P4-IPsec}
We first give an overview on the P4 processing pipeline of P4-IPsec.
For the sake of simplicity in presentation, we combine functions into function blocks and describe them in detail.

\subsubsection{P4 Processing Pipeline}
\fig{data-plane-overview} depicts the P4 packet processing pipeline of P4-IPsec.
It consists of a parser, deparser, and four function blocks in between.
When a packet arrives via the ingress, the P4 parser first extracts the packet headers.
In case of a header other than \ac{ESP}, the parser forwards the packet to optional higher-layer functions that operate on protocol layers such as TCP or UDP.
Afterwards, the \ac{SPD} matching function block processes the packet.
Following the IPsec standard, entries in the \ac{SPD} determine about the action to be executed on the packet.
In case of DISCARD, the packet is dropped.
In case of BYPASS, the packet is passed to the L3 forwarding function block.
In case of PROTECT, the packet is passed to the ESP encrypt function block.
In the ESP encrypt control block, encryption using \ac{SA} data from the SAD\textsubscript{enc} \ac{MAT} is applied to the IP packet.
In the L3 forwarding control block, the packet is forwarded based on rules defined in the forwarding \ac{MAT}.
Going back to the parser again: if the packet has an ESP header, it is forwarded to the ESP decryption control block.
It validates the packet's authenticity, decrypts the ESP message, and extracts the original IP packet that is then passed to the L3 forwarding control block.
In case of missing entries in the SPD, SAD\textsubscript{dec}, SAD\textsubscript{enc}, or LPM-FWD \ac{MAT}, the packet is dropped.
As final step, the deparser reassembles all headers and re-calculates the IPv4 checksum as some fields, e.g., the TTL, are changed.
Runtime behavior of the data plane can be managed by manipulating the \acp{MAT} via an SDN controller.

\subsubsection{Function Block: L3 Forwarding}
\begin{figure}[ht]
    \begin{center}
    \includegraphics[width=.99\linewidth]{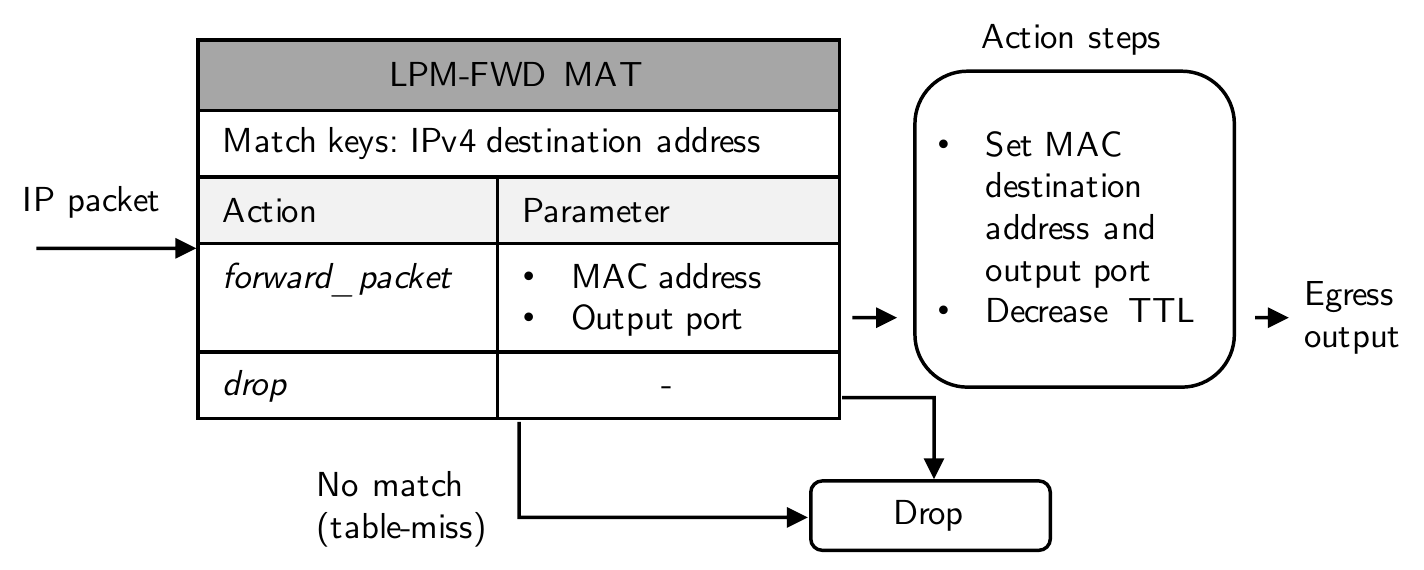}
    \end{center}
    \caption{L3 forwarding function block. IP packets are processed by the LPM-FWD MAT that either applies the forward\_packet or drop action. In case of no match, the drop action is applied.}
    \label{fig:forward-control-block}
\end{figure}

\fig{forward-control-block} depicts the function block of \textit{L3 forwarding}.
It implements packet forwarding to the next hop via a particular output port of the P4 switch.
The LPM-FWD \ac{MAT} matches packets using their IPv4 destination addresses to two actions: \textit{forward\_packet} and \textit{drop}. 
The forward\_packet action receives the MAC address of the next hop and the output port as parameters from the \ac{MAT}.
Then, it sets the MAC destination address of the packet to the MAC address of the next hop, decreases the \ac{TTL} by 1 in the IP header, and sets the output port.
Afterwards, the packet is forwarded to the deparser and sent out via the egress.
\textit{drop} directly discards the packet; this action is also applied if no match in the LPM-FWD \ac{MAT} is found.

\subsubsection{Function Block: SPD Matching}
\begin{figure}[ht]
    \begin{center}
    \includegraphics[width=.98\linewidth]{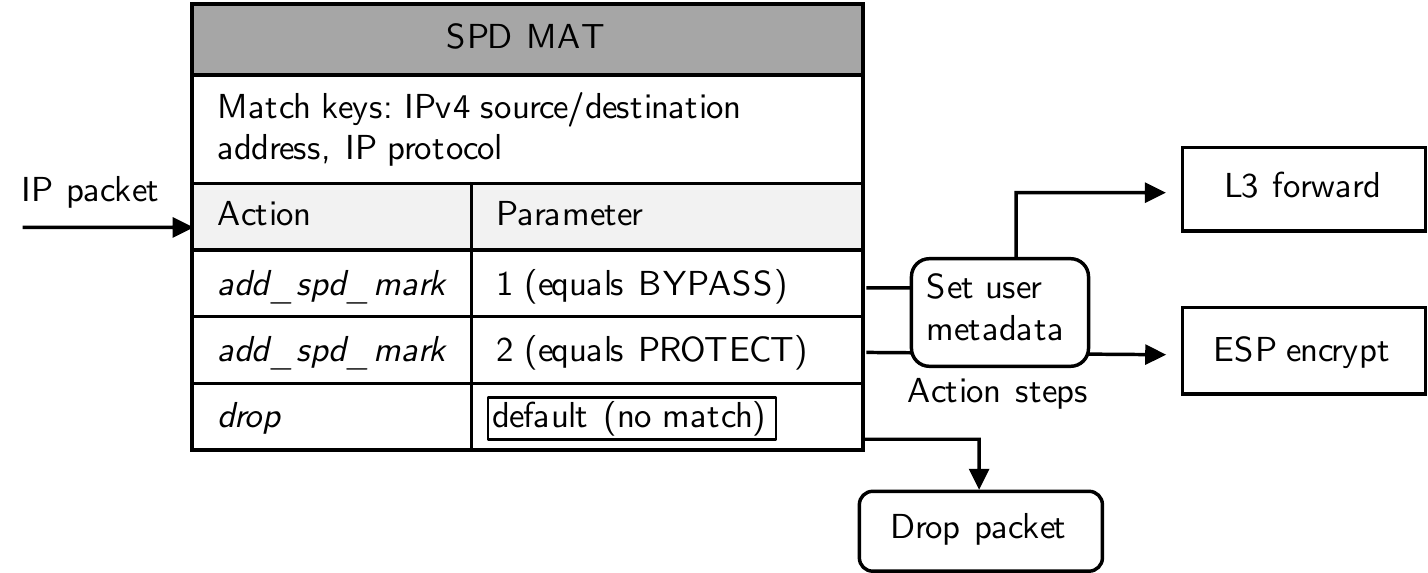}
    \end{center}
    \caption{SPD matching function block. IP packets are processed by the SPD MAT. The add\_spd\_mark action adds the given parameter to the user metadata of the packet. It decides if the packet is protected by IPsec (PROTECT) or forwarded without protection (BYPASS) in later stages.}
    \label{fig:spd-matching-control-block}
\end{figure}

\fig{spd-matching-control-block} depicts the function block of \textit{SPD matching}.
We introduce a \ac{SP} \ac{MAT} that resembles the \ac{SPD} from the IPsec standard  (see \sect{ipsec-foundations}).
It matches given packets with \ac{SPD} rules and adds a mark to the user metadata of each packet that is used in further processing within the P4 processing pipeline.
We implement IPv4 source and destination address and IP protocol as exemplary match keys.
Due to P4's flexibility in defining packet parsers and parsing packets, more match keys, e.g., for TCP/UDP ports or even application-layer ports could be added easily.
Actions are either \textit{add\_spd\_mark} or \textit{drop}.
The \textit{add\_spd\_mark} action adds "spd\_mark = 1" for BYPASS or "spd\_mark = 2" for PROTECT to the user metadata field of the packet.
\textit{drop} directly discards the packet; this action is also applied if no match is found.

\subsubsection{Function Block: ESP Encryption}
\begin{figure}[ht]
    \begin{center}
    \includegraphics[width=.99\linewidth]{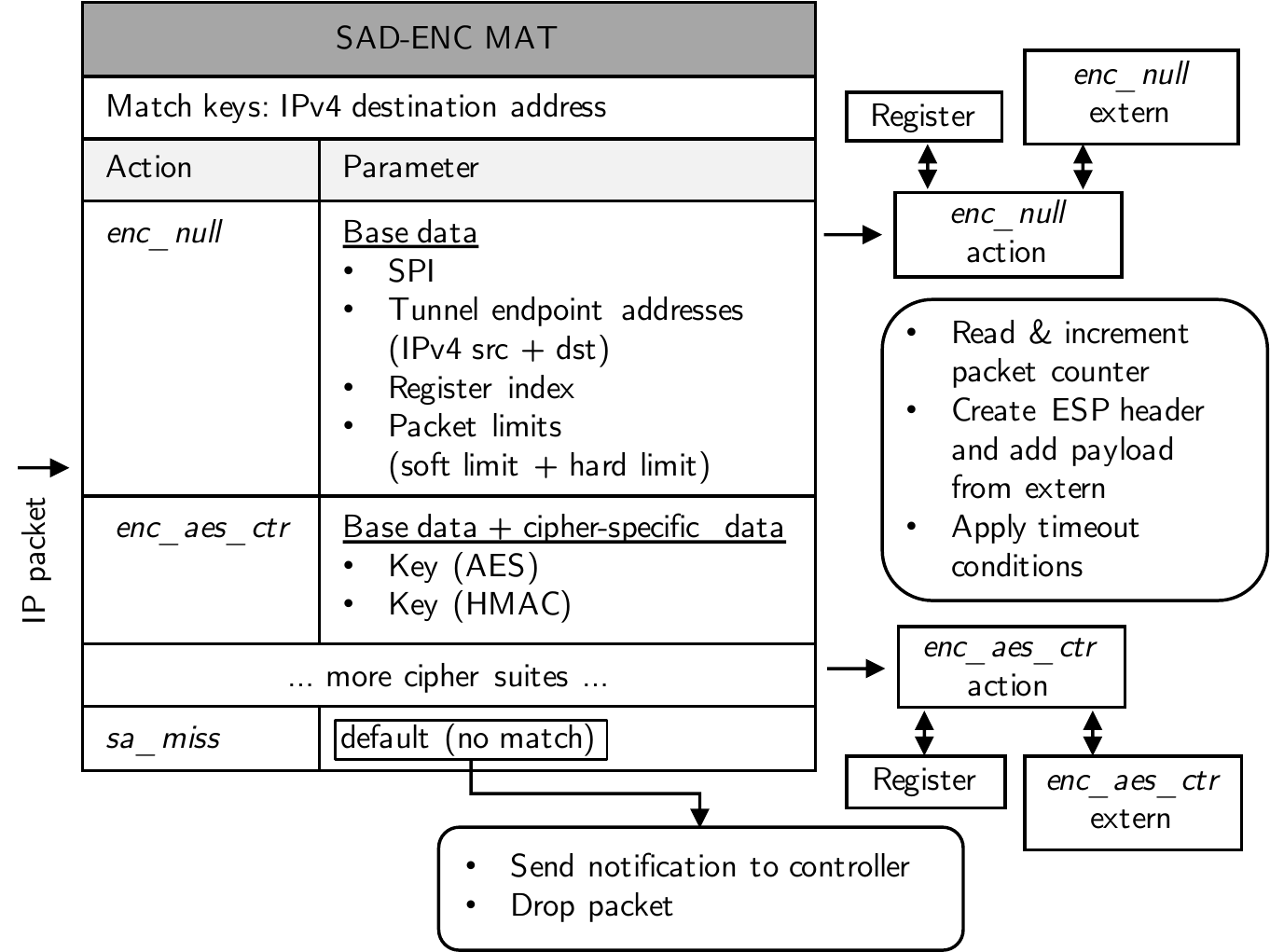}
    \end{center}
    \caption{ESP encryption function block. IP packets are processed by the SAD-ENC MAT. It holds entries for each SA with the corresponding data that is required for applying the associated cipher suite externs for encryption.}
    \label{fig:esp-encrypt-function-block}
\end{figure}

\fig{esp-encrypt-function-block} depicts the function block of \textit{ESP encryption}.
We introduce a SAD-ENC \ac{MAT} that resembles the \ac{SAD} from the IPsec standard (see \sect{ipsec-foundations}).
Each entry in the \ac{MAT} represents a particular \ac{SA} that is identified by the IPv4 destination address, i.e., packets are matched based on their IPv4 destination address.

We implement cipher suites as actions that rely on externs and registers.
Representing a variety of cipher suites, we implement the \textit{AES-CTR} and \textit{NULL} cipher suite as examples.
The \textit{NULL} cipher suite is intended for testing purposes only.
It uses the identity function instead of encrypting data and skips calculating an integrity check value.
Cipher suite actions receive two types of parameters: basic parameters that are required by all cipher suites and cipher-specific parameters.

We first describe the basic parameters.
The \textit{\ac{SPI}} is part of the \ac{ESP} header.
It identifies the \ac{SA}.
The \textit{tunnel endpoint addresses} (IPv4 source/destination address) identify the source and destination of the IPsec tunnel.
Both are part of the new outer IPv4 header that encapsulates the \ac{ESP} frame.
The \textit{register index} points to a particular index that holds the packet counter for the particular \ac{SA} used by the cipher suite extern.
\textit{Packet limits} declare timeout conditions in terms of packet count thresholds for \acp{SA}.
If a soft limit is reached, rekeying is triggered.
If a hard limit is reached, packets that belong to that \ac{SA} are dropped.

The NULL cipher suite is an example that only requires this set of basic parameters.
Typical cipher suites that implement particular encryption and authentication mechanisms require additional parameters such as keys, \acp{IV}, or even additional constructs to keep cipher state, e.g., registers.
AES-CTR, as example for such a cipher suite, requires a key for AES and a key for HMAC.

The functionality within the cipher suite action is as follows.
First, the packet counter for the particular \ac{SA} is read from the register and incremented.
Second, an \ac{ESP} header is created with the \ac{SPI} and sequence number of the packet.
For the creation of the \ac{ESP} packet, the action passes the original \ac{IP} packet, the newly created \ac{ESP} header, and required keys of the cipher suite to the corresponding extern.
The cipher suite extern performs encryption/authentication and responds with the \ac{ESP} packet.
Fourth, the new outer \ac{IP} packet is created with the tunnel endpoint addresses.
It encapsulates the newly created \ac{ESP} packet.
Last, timeout conditions are checked.
The user metadata structure includes flags for \textit{soft\_limit\_reached} and \textit{hard\_limit\_reached} that are set in case of matching conditions.

\subsubsection{Function Block: ESP Decryption}
\begin{figure}[ht]
    \begin{center}
    \includegraphics[width=.99\linewidth]{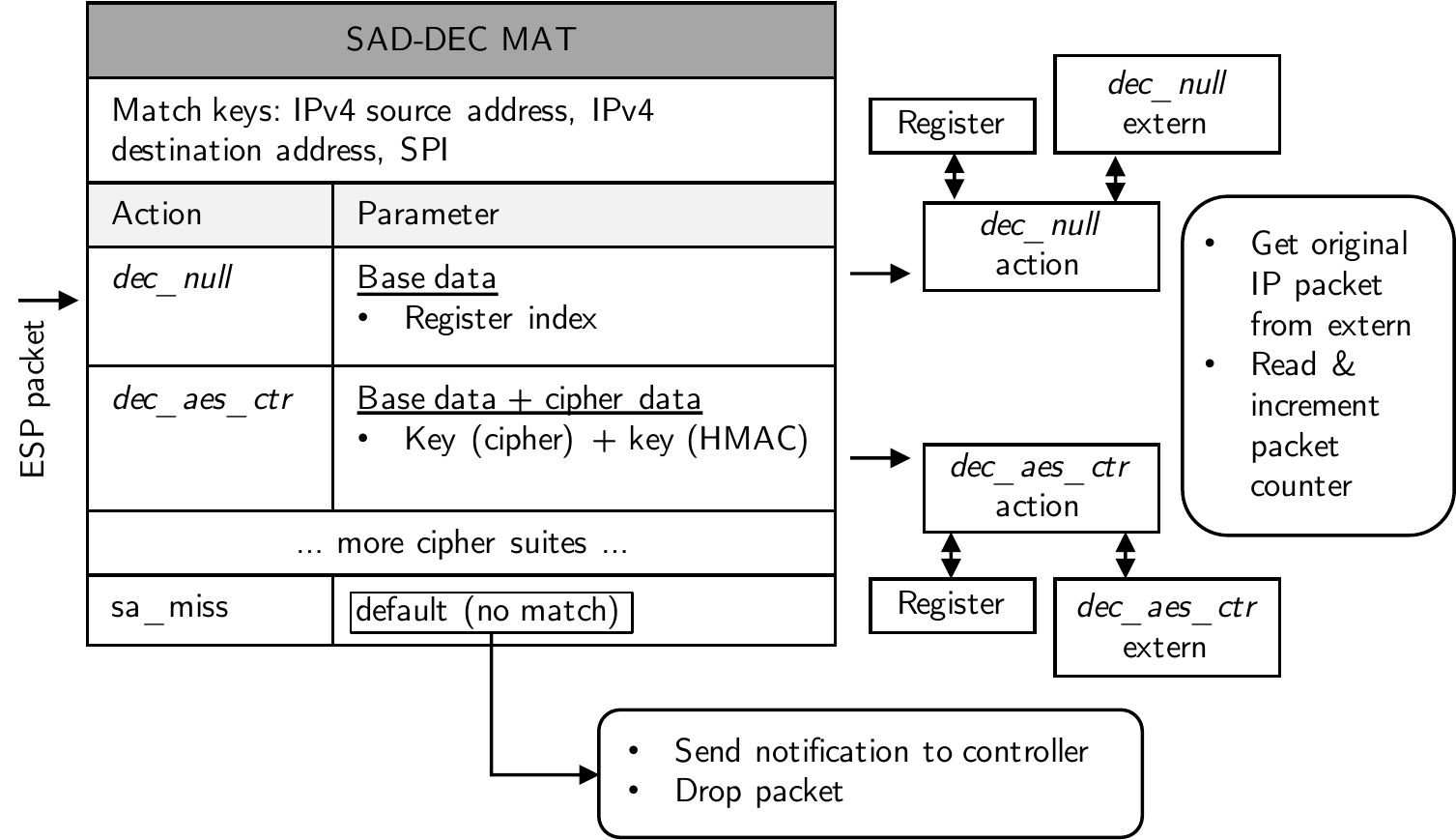}
    \end{center}
    \caption{ESP decryption function block. ESP packets are processed by a SAD-DEC MAT. It holds entries for each SA with the corresponding data that is required for applying the associated cipher suite externs for decryption.}
    \label{fig:esp-decrypt-function-block}
\end{figure}

\fig{esp-decrypt-function-block} depicts the function block of \textit{ESP decryption}.
We introduce the SAD-DEC \ac{MAT} that resembles the decryption \ac{SAD} from the IPsec standard (see \sect{ipsec-foundations}).
Each entry in the \ac{MAT} represents a particular \ac{SA} that is identified by the outer IPv4 source address and IPv4 destination address (tunnel endpoints), and the \ac{SPI}.
As in the function block of ESP Encryption, cipher suites are implemented as actions that rely on externs and registers with a different set of action parameters. 

The functionality within the cipher suite action is as follows.
First, the packet counter for the particular \ac{SA} is read from the register and incremented.
Second, the original \ac{IP} packet is extracted from the \ac{ESP} packet.
Therefore, the action passes the \ac{ESP} packet and the required keys of the cipher suite to the corresponding extern.
The cipher suite extern performs decryption/authentication and responds with the original \ac{IP} packet.
Last, timeout conditions are checked as described in ESP encryption.

\subsection{Control Plane Operation of P4-IPsec}
\label{sec:concept-dns}

We first give an overview on the control plane operation of P4-IPsec.
We describe how configuration data is generated on the controller and how it is set up in both, host-to-site and site-to-site operation mode.

\subsubsection{Overview}
\begin{figure}[ht]
    \begin{center}
    \includegraphics[width=.9\linewidth]{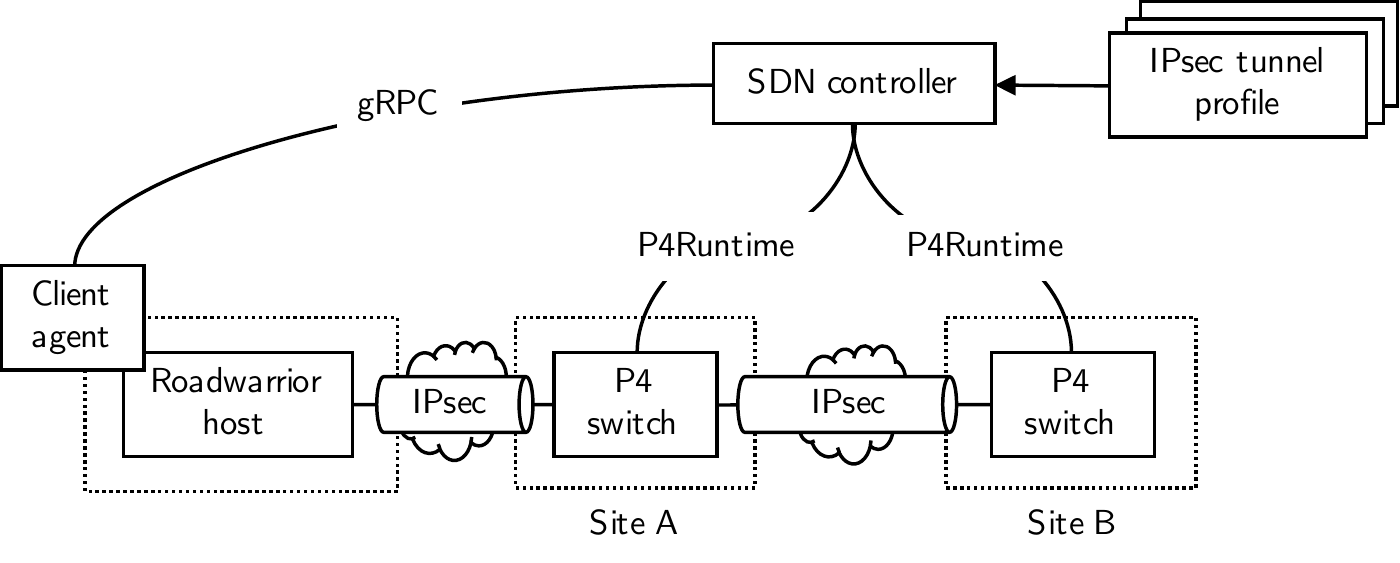}
    \end{center}
    \caption{Control plane operation in P4-IPsec. The client agent on the roadwarrior host holds a control channel via gRPC to the SDN controller. P4 switches are connected via P4Runtime. The SDN controller includes IPsec tunnel profiles with configuration data for tunnel setup and management.}
    \label{fig:control-plane-overview}
\end{figure}

\fig{control-plane-overview} depicts the control plane interaction in the two operation modes of P4-IPsec, host-to-site and site-to-site mode.
In both operation modes, IPsec tunnels are set up by the controller on the basis of IPsec tunnel profiles.
Those can be manually defined by an administrator or generated by another software component, e.g., a network operation platform.
In host-to-site operation mode, the SDN controller interacts with the client agent via a gRPC tunnel and with the P4 switch via P4Runtime.
In site-to-site operation mode, the SDN controller interacts with both P4 switches via P4Runtime.
On roadwarrior hosts, configuration data is converted into \emph{ip xfrm} commands that set up the tunnel.
For P4 switches, the controller directly writes to \acp{MAT} and receives notifications, e.g., if an \ac{SA} needs to be renewed, via P4Runtime.
For the sake of simplicity, we restrict our implementation to proactive IPsec tunnel setup.
In site-to-site mode, \ac{IPsec} tunnels are set up and kept alive for all configured P4 switches.
For host-to-site mode, the client agent presents a selection of available tunnels.
The user then can select one or multiple \ac{IPsec} tunnel profiles to be set up by the controller.

This mechanism can be extended or substituted by more sophisticated approaches such as on-demand VPN setup.
Pre-defined conditions (e.g., a request for a network resource in an internal network) may trigger \ac{IPsec} tunnel setup via the controller.

\subsubsection{IPsec Tunnel Profiles}
An IPsec tunnel between two peers consists of two unidirectional \acp{SA}, each identified by a unique \ac{SPI}.
Due to their direction, the first peer of an \ac{SA} is called "left" where the second peer of the \ac{SA} is called "right".
We denote the \ac{SA} from the left to the right peer as \ac{SPI}\textsubscript{i} and the \ac{SA} from the right to the left peer as \ac{SPI}\textsubscript{j}, respectively.
Each \ac{SA} requires two \ac{MAT} entries: one for encrypting \ac{ESP} packets in the SAD-ENC \ac{MAT} and one for decrypting \ac{ESP} packets in the SAD-DEC \ac{MAT}.

\begin{figure}[ht]
    \begin{center}
    \includegraphics[width=.99\linewidth]{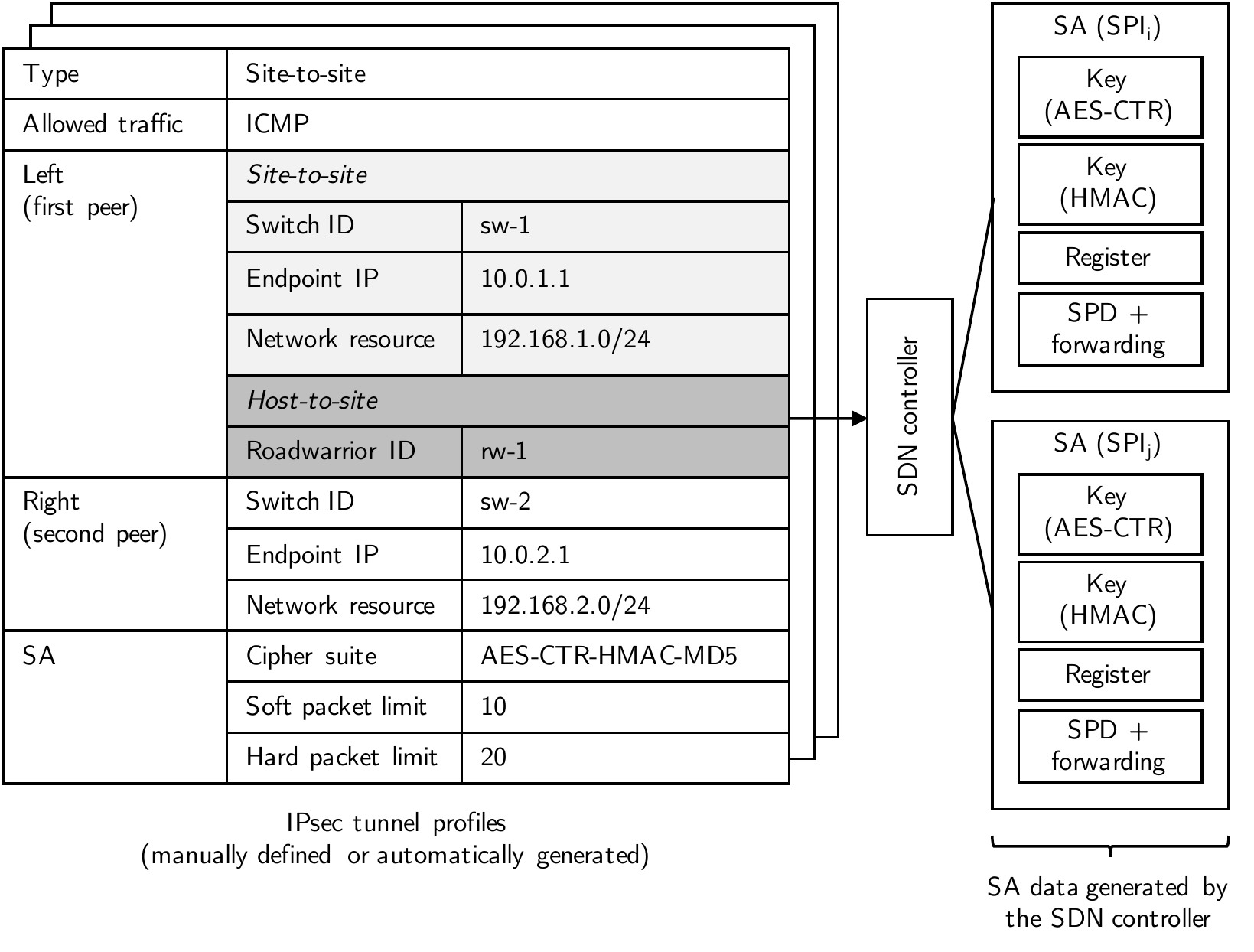}
    \end{center}
    \caption{IPsec tunnel profiles with the associated SA data generated by the SDN controller. Configuration data in the IPsec tunnel profiles depends on the operation mode (host-to-site or site-to-site). SA data depends on the cipher suite that is defined in the IPsec tunnel profile.}
    \label{fig:ipsec-tunnel-profile}
\end{figure}

\fig{ipsec-tunnel-profile} depicts how the SDN controller generates configuration data for the roadwarrior hosts or P4 switches.
IPsec tunnel profiles are the basis for any IPsec tunnel.
As basic information about the tunnel, it includes information about the type of IPsec tunnel (host-to-site or site-to-site) and the allowed traffic that is set to PROTECTED via \ac{SPD} rules.
The left peer (first) can be a P4 switch (in site-to-site operation mode) or a roadwarrior host (in host-to-site operation mode).
In case of site-to-site operation mode, this field holds the switch ID (unique identifier of the P4 switch), endpoint IP (public IP address of the P4 switch), and network resource (internal network behind the P4 switch).
In case of host-to-site operation mode, this field only holds the roadwarrior ID.
The right peer (second) is always a P4 switch.
Therefore, it holds the same data as in the left peer field in site-to-site operation mode as described before.
The SA field holds the cipher suite and soft/hard packet limits.
On the basis of an IPsec tunnel profile, the controller generates configuration data for both \acp{SA}.
In case of the AES-CTR-HMAC-MD5 cipher suite, \ac{SA} data inludes keys for AES-CTR and HMAC, register indexes, and configuration data for the SPD and forwarding function block.
In case of the NULL cipher suite, keying material is not needed.

In our prototype, IPsec tunnel profiles are manually defined by an administrator.
In practice, they can be generated by a software component, e.g., a network operation platform, on the basis of user/device profiles, groups, network resources, and permission models.

\subsubsection{Controller Connection}
We describe the management connections in site-to-site and host-to-site operation mode.

\paragraph{Site-to-Site (P4Runtime)}
Site-to-site mode relies on P4Runtime for managing the P4 switches.
Explained in Section \ref{foundations:p4-runtime}, the control plane connection to the P4 switches is established by the SDN controller.
Therefore, it holds a list of connection data (name, address, port identity) of all assigned P4 switches.

\begin{figure}[ht]
    \begin{center}
    \includegraphics[width=.9\linewidth]{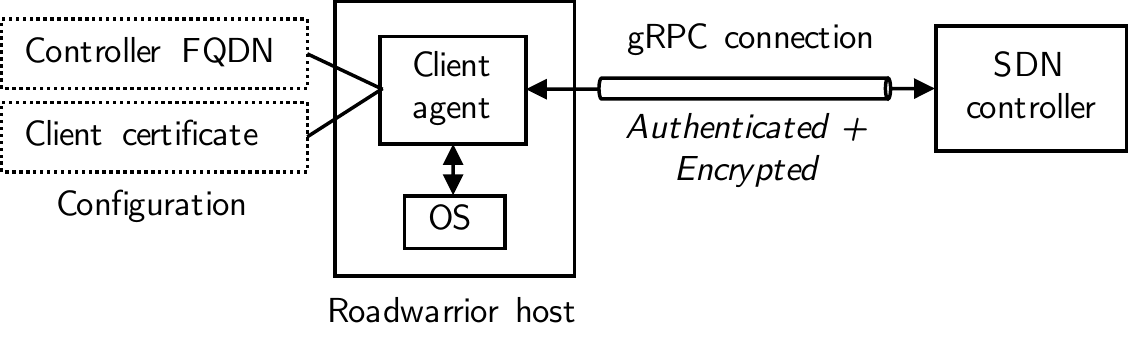}
    \end{center}
    \caption{Control plane connection between the client agent and SDN controller. The client agent depends on the FQDN of the controller and a client certificate to establish a gRPC connection to the SDN controller.}
    \label{fig:control-plane-connection-host-to-site}
\end{figure}

\paragraph{Host-to-Site (gRPC)}
\fig{control-plane-connection-host-to-site} depicts the connection between the client agent running on the roadwarrior host and the SDN controller.
Required configuration data for the start of the client agent are the FQDN of the controller and the client certificate.
At start, the client agents establishes a gRPC tunnel to the SDN controller.
The gRPC tunnel is protected with SSL/TLS, i.e., the client agent and SDN controller perform a mutual authentication using certificates and establish an encrypted connection.
Certificates can be created and deployed to all roadwarrior hosts running the client agent and the SDN controller with a \ac{PKI}.
Roadwarrior host access can be removed by simply revoking the associated client certificate.
In addition, gRPC provides support for optional multi-factor authentication (MFA) with token-based authentication via the Google Authenticator service.
After connection setup, the client agent and SDN controller exchange configuration and signaling data via the gRPC tunnel.
The client agent implements interfaces to interact with the roadwarrior host's operating system for configuration and signaling.
Control plane connection to the P4 switch as remote peer is established with P4Runtime as described before.

Host-to-site operation mode requires that the controller is dual-homed.
It has an interface to a management network where it holds P4Runtime connections to the P4 switches and another interface that makes it accessible via the Internet for client agents running on roadwarrior hosts.
On the latter interface, it has an IP address that is publically reachable via the Internet.
Although mutual certificate-based authentication protects against malicious P4-IPsec agents, this public interface should be protected as every publically available web service, e.g., with a firewall.

\subsubsection{Tunnel Management Operations}
\label{sec:tunnel-management}
We describe the elementary operations of tunnel setup, tunnel renewal, and tunnel deletion that apply to both operation modes.

\begin{figure}[ht]
    \begin{center}
    \includegraphics[width=.9\linewidth]{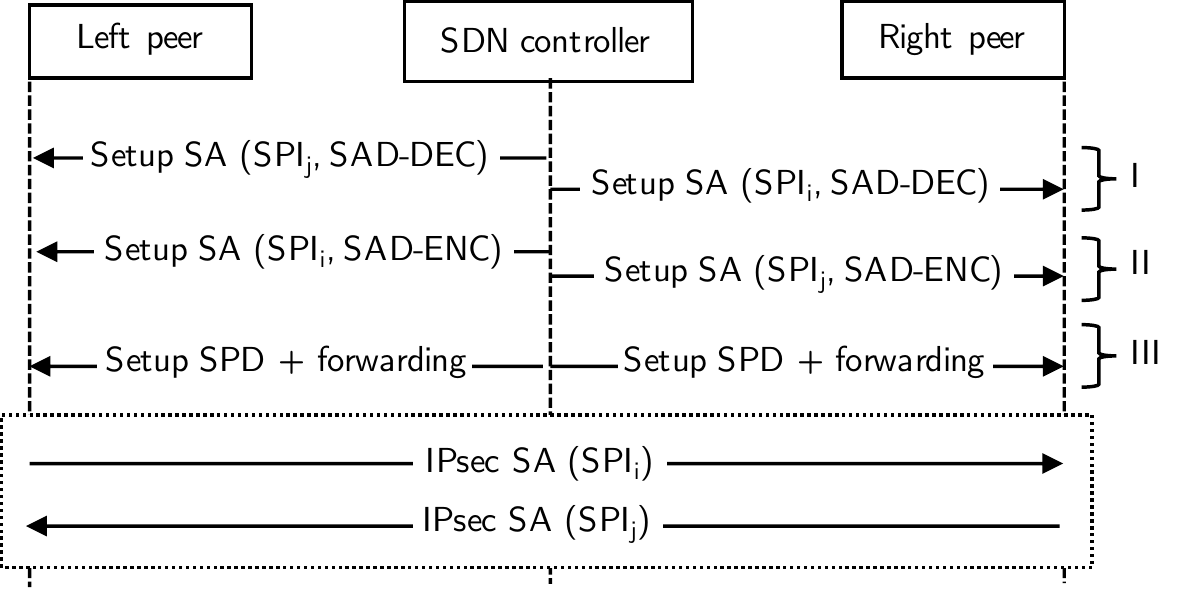}
    \end{center}
    \caption{Setup procedure for a bidirectional IPsec tunnel on two peers. The SDN controller sets up decryption and encryption for both SAs (II) followed by SPD and forwarding entries.}
    \label{fig:client-tunnel-setup}
\end{figure}

\paragraph{Tunnel Setup}
\fig{client-tunnel-setup} depicts the three-step process of IPsec tunnel setup.
First, the controller sets up both \acp{SA} in the SAD-DEC \acp{MAT} of both peers.
Second, the controller sets up both \acp{SA} in the SAD-ENC \acp{MAT} of both peers.
Setting up \ac{SA} entries for decryption first ensures that no \ac{ESP} packets get lost if one peer immediately starts to send \ac{ESP} packets.
Last, the controller sets up the \ac{SPD} and installs forwarding rules if required.

\begin{figure}[ht]
    \begin{center}
    \includegraphics[width=.99\linewidth]{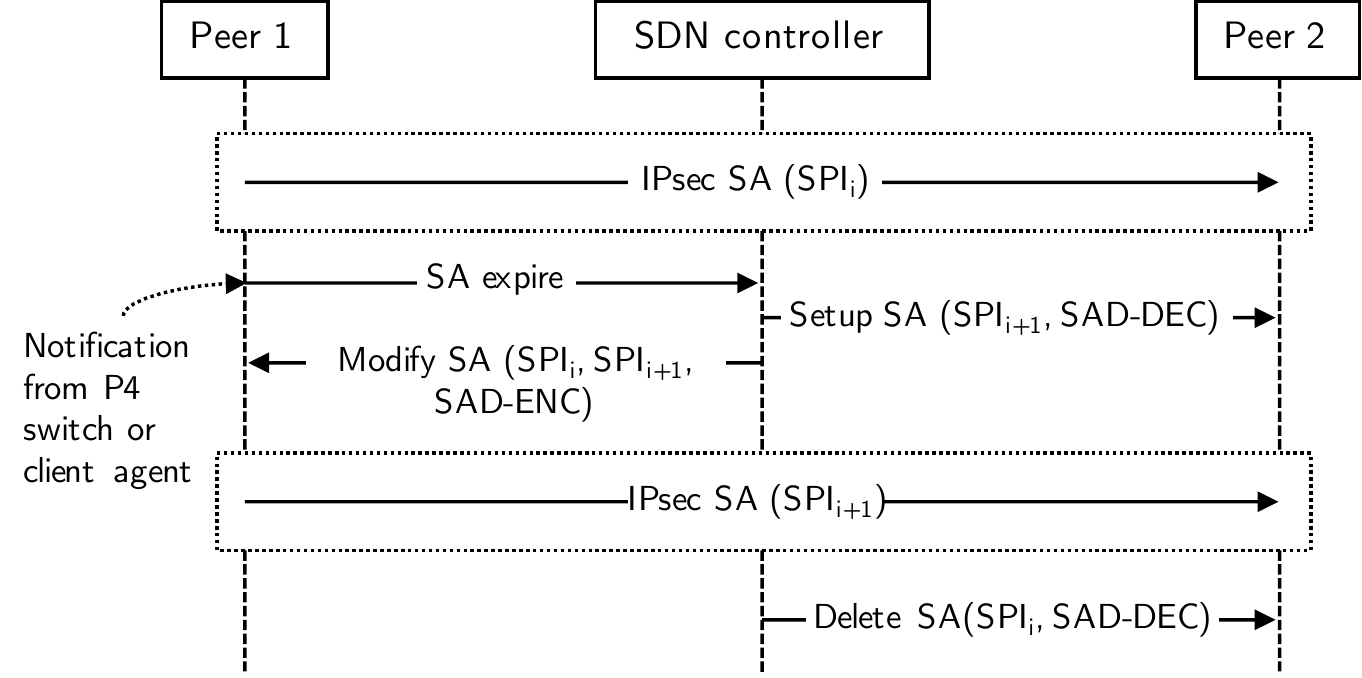}
    \end{center}
    \caption{Renewal procedure for an unidirectional SA within an IPsec tunnel on two peers. After receiving an SA expire message, the SDN controller installs a new decryption SA on the remote peer. Afterwards, it replaces the expired encryption SA with new SA data. As cleanup step, the old decryption SA is removed.}
    \label{fig:client-tunnel-renew}
\end{figure}

\paragraph{Tunnel Renewal}
IPsec SAs have a limited lifetime, i.e., keying material needs to be renewed on a regular basis.
Both, client agent and P4 switch notify the SDN controller if an \ac{SA} needs to be renewed.
For the client agent, this notification is triggered by the kernel implementation of IPsec that sends expiration messages in the case that soft and hard timeout limits are reached.
For the P4 switches, this is implemented using packet counters in registers that are checked with each packet processing within an ESP encrypt or decrypt function block.
We adopt the principle presented by Lopez-Millan \etal{} \cite{LoMa19} that implements tunnel renewal without risking packet loss.
\fig{client-tunnel-renew} depicts the process of IPsec tunnel renewal.
When the SDN controller received the \ac{SA} expire notification, it generates a new \ac{SA} that is identified by a new \ac{SPI}.
Then, tunnel renewal follows the principle of tunnel setup as described before.
First, the new \ac{SA} is installed in the SAD-DEC \ac{MAT}.
Second, the existing \ac{SA} in the SAD-ENC \ac{MAT} is replaced by the new \ac{SA} via a modify operation.
Last, as it can be ensured that no packets are encrypted using the previous \ac{SA}, its entry can be removed from the SAD-DEC \ac{MAT}.

\paragraph{Tunnel Deletion}
If an IPsec tunnel should be deleted, the SDN controller removes the associated entries in the SPD, SAD-ENC, and SAD-DEC \ac{MAT}.
This is triggered on the SDN controller, e.g., when an IPsec profile is removed.
\section{Prototypical Implementation}
\label{sec:implementation-mininet}

We describe our software-based prototype of P4-IPsec.
We outline its three parts, the P4 data plane, the client agent running on the roadwarrior host, and the control plane implementation, in detail.
We publish the implementation and our testbed environment under the Apache v2 license on GitHub \cite{p4-ipsec-github}.

\subsection{Testbed Environment}
\label{sec:testbed-environment}

Our prototypical implementation of P4-IPsec includes a softwarized testbed environment.
We use Mininet to create network topologies that consist of \ac{BMv2} P4 switches and network hosts.
We build it with Vagrant \cite{vagrant}, a tool that simplifies creation and management of virtual environments.
All resources and setup steps are part of a configuration file.
Executing \textit{vagrant up} in a console within the repository folder automatically sets up and launches the testbed environment.
The testbed environment includes a virtual machine running Ubuntu 16.04 with all dependencies: libyang, sysrepo, mininet, protobuf, gRPC, PI/P4Runtime, BMv2, and P4C.
The versions of all components can be found in the setup scripts.

\subsection{Data Plane Implementation}
We implement the P4 data plane implementation for the \ac{BMv2} P4 software target.
We extend its \textit{simple\_switch} architecture by externs programmed in C++ for the AES-CTR-HMAC-MD5 and NULL \ac{IPsec} cipher suites.
Each cipher suite is implemented by two externs, one for encryption and one for decryption.
For AES-CTR-HMAC-MD5, we use OpenSSL to apply AES-CTR for encryption/decryption and HMAC-MD5 for packet authentication.
We implement the P4 processing pipeline as P4\textsubscript{16} program.
It relies on the cipher suite externs and uses registers to store packet counters for the \acp{SA}.
We run the P4 program on our extended \textit{simple\_switch} P4 target.
We encapsulate our modified simple\_switch P4 target within the \textit{simple\_switch\_grpc} P4 target so that P4Runtime API can be used for interaction with the \ac{SDN} controller.

\subsection{Client Agent}
We implement the client agent as Python 3.6 command line tool for Linux hosts.
We integrate a gRPC client using the gRPC library \cite{grpc} as interface to the \ac{SDN} controller.
For \ac{IPsec} tunnel setup, the client agent translates configuration data from the \ac{SDN} controller into particular \textit{XFRM} commands from the \textit{iproute2} tool to configure \ac{IPsec} on the roadwarrior host.
In addition, it sets up IP routes for routing IP traffic via the \ac{IPsec} tunnel.
Received and applied configuration data is cached so that proper teardown configuration can be applied in case of tunnel shutdown.
We implement rekeying with the help of \textit{Netlink} \cite{netlink}.
The client agent monitors Netlink messages by listening on the corresponding Netlink socket and binding to the XFRMNLGRP\_EXPIRE address so that XFRM Expire messages can be received.
When receiving an XFRM Expire message, it extracts parameters such as \ac{SPI} and IP addresses of the tunnel endpoints.
To initiate rekeying, the tunnel source and destination address and \ac{SPI} are put into a queue for processing in the main class.

\subsection{SDN Controller}
We implement the controller as command line tool in Python 2.7.
We use the p4runtime\_lib \cite{p4-runtime-lib} to integrate the interface to the P4 switch and the gRPC library to integrate the interface to the client agent.
The controller features a simple \ac{CLI} for development and testing purposes that displays information about all active \ac{IPsec} tunnels.
P4Runtime and p4runtime\_lib facilitate easy implementation of individual controllers for prototypes.
Nevertheless, those functions could be also integrated into existing \ac{SDN} controllers such as ONOS or OpenDaylight.
\section{Performance Evaluation with the Software Switch BMv2}
\label{sec:evaluation-mininet}

We describe the test environment and report experiment results performed with the P4-IPsec software prototype introduced in \sect{implementation-mininet}.

\subsection{Methodology}
We conduct the performance experiments in the testbed environment presented in \sect{testbed-environment}.
The testbed runs on a Lenovo Thinkpad T480s (Intel i5-8250U CPU, \SI{16}{\giga\byte} RAM) with Manjaro Linux.
The Vagrant file in the repository includes the version numbers of all software components from the testbed environment.

\begin{figure}[ht]
    \begin{center}
    \includegraphics[width=.8\linewidth]{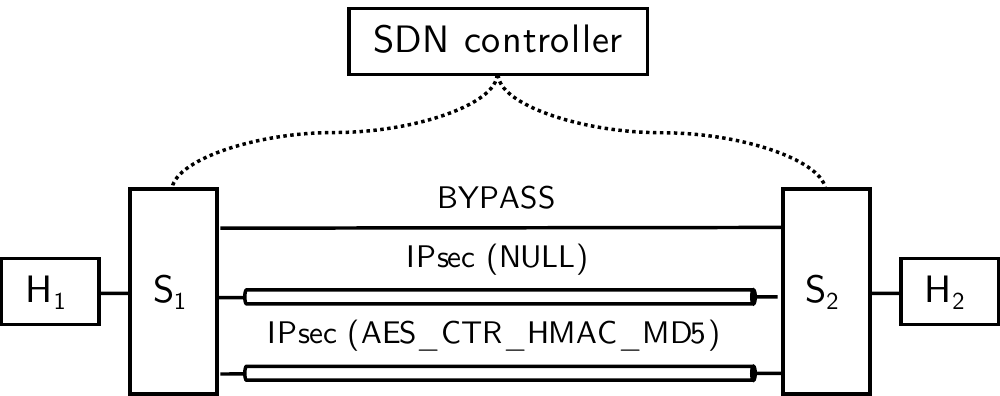}
    \end{center}
    \caption{Experiment setup for evaluation of the P4-IPsec prototype on the BMv2 switches S\textsubscript{1} and S\textsubscript{2}.}
    \label{fig:experiment-setup}
\end{figure}

\fig{experiment-setup} depicts the experiment setup.
S\textsubscript{1} and S\textsubscript{2} are \ac{BMv2} P4 switches,  H\textsubscript{1} and H\textsubscript{2} are Linux hosts that are attached to them.
S\textsubscript{1} and S\textsubscript{2} are connected via two unidirectional virtual links.
We do not configure any additional delay or bandwidth limitations on these links.
Traffic between H\textsubscript{1} and H\textsubscript{2} is forwarded by S\textsubscript{1} and S\textsubscript{2}.
We set up IPsec tunnels with different cipher suites and conduct TCP goodput measurements between H\textsubscript{1} and H\textsubscript{2} using iperf in version 3.7.

\begin{figure*} 
    \hfill
    \subfigure[TCP goodput measured by iperf3 on the receiving host. Three variants are considered: without IPsec forwarding (BYPASS), with IPsec forwarding but without encryption, and with IPsec forwarding using the AES\_CTR\_HMAC\_MD5 cipher suite.\label{fig:goodput}]{\includegraphics[width=0.32\linewidth]{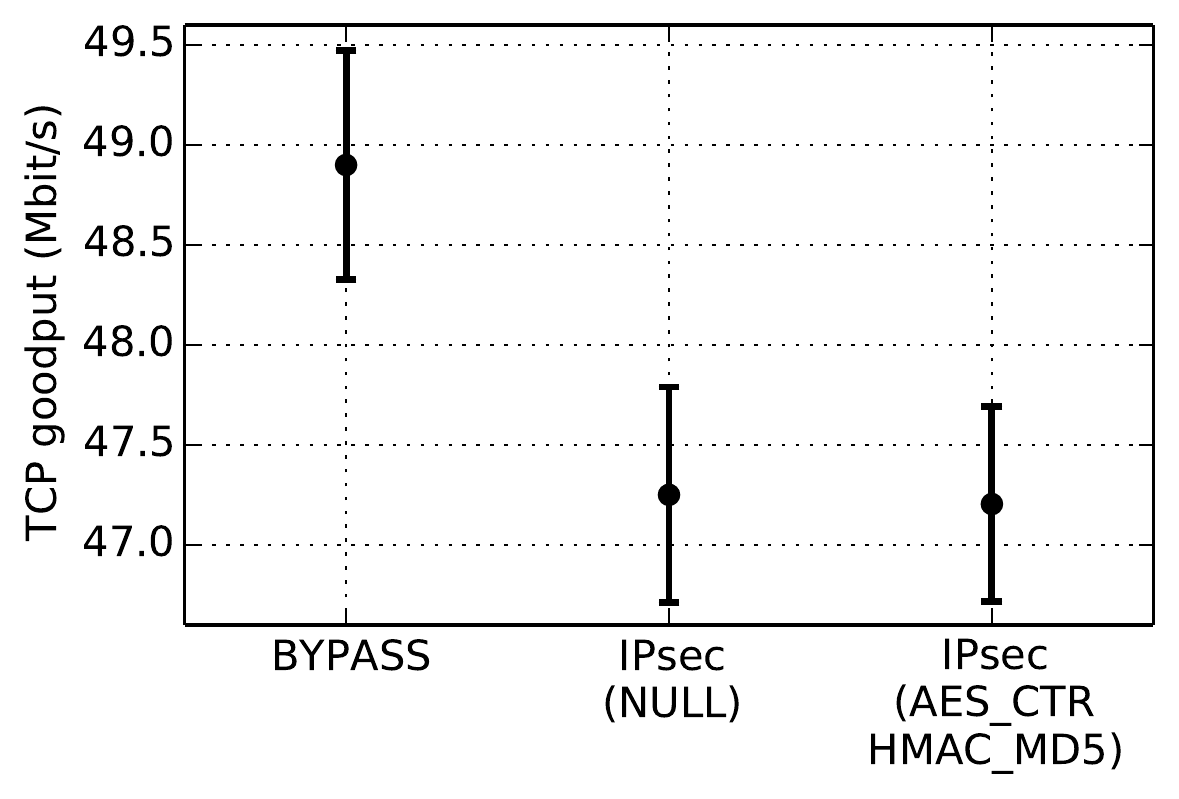}}
    \hfill
    \subfigure[IPsec tunnel setup and renewal times. Times are measured on the
    controller from initiation to completion of these processes.\label{fig:setup-renewal-times}]{\includegraphics[width=0.32\linewidth]{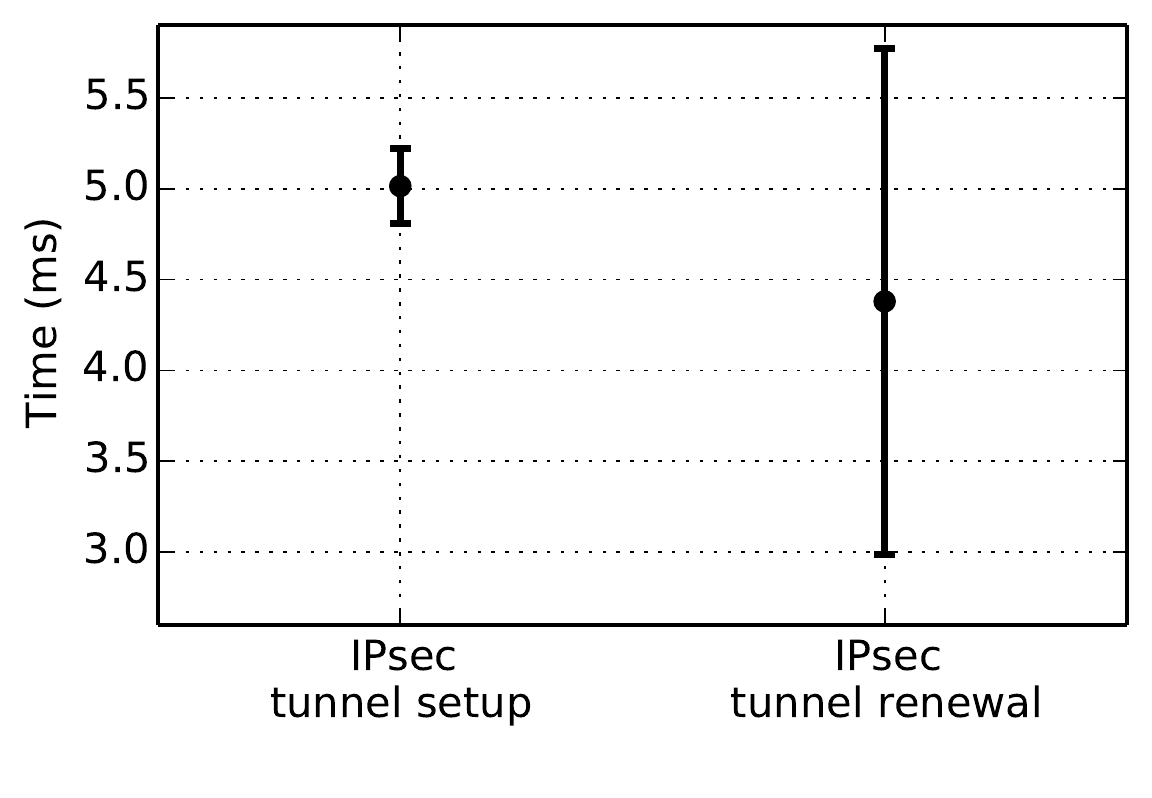}}
    \hfill
    \subfigure[Times for creation of a single \ac{SA}, insertion of a single \ac{MAT} entry, and update of a single \ac{MAT} entry. Times are measured on the controller from initiation to completion of these processes.\label{fig:basic-operation-times}]{\includegraphics[width=0.32\linewidth]{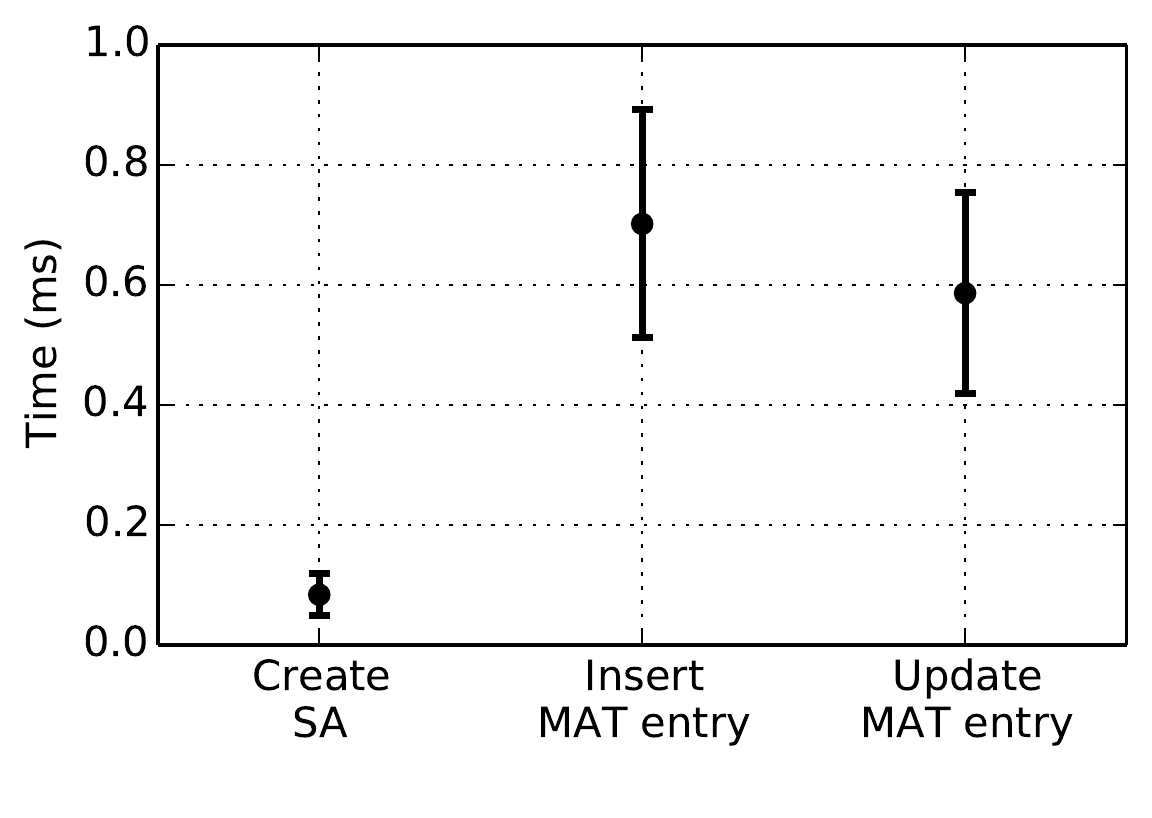}}
    \hfill
    \caption{Measurement results for experiments with P4-IPsec using \ac{BMv2} software switches in a virtual environment with almost zero link delays. Average values are shown with confidence intervals for a significance of $\alpha = 5\%$.}
\end{figure*}

The results in the following represent measured average values with confidence intervals for a significance of $\alpha = 5\%$.
Thus, the true averages lie within the displayed ranges with a probability of $1-\alpha$.

\subsection{Data Plane Evaluation}
We investigate how P4-IPsec's data plane implementation affects the overall throughput on the \ac{BMv2} software target.

\subsubsection{Experiment Description}
We analyze TCP goodput for three different configurations.
In the first scenario, we install BYPASS rules in the \ac{SPD} so that traffic is only forwarded and not handled by IPsec.
In the second scenario, we establish an IPsec tunnel between S\textsubscript{1} and S\textsubscript{2} with the NULL cipher suite.
In the third scenario, we establish an IPsec tunnel between S\textsubscript{1} and S\textsubscript{2} with the AES\_CTR\_HMAC\_MD5 cipher suite.
Each experiment comprises 20 runs, each with a duration of \SI{60}{\s} and a MTU set to \SI{1450}{\byte}.
We configure soft and hard packet limits for rekeying to $50000$ and $51000$ resulting in an average of six rekeyings per run, three for each of the two \acp{SA} of the IPsec tunnel.

\subsubsection{Results \& Discussion}
\fig{goodput} depicts the results.
For forwarding without IPsec (BYPASS), TCP goodput is \SI{48.90}{\mega\bit\per\s}.
For IPsec forwarding with the NULL cipher suite, TCP goodput is \SI{47.25}{\mega\bit\per\s}.
For IPsec forwarding with the AES\_CTR\_HMAC\_MD5 cipher suite, TCP goodput is \SI{47.21}{\mega\bit\per\s}.
Hence, the experimental results show similar TCP goodput rates between \SI{47.21}{\mega\bit\per\s} and \SI{48.90}{\mega\bit\per\s} for all three scenarios.
The drop in performance is caused by IPsec processing while the differences between both IPsec cipher suites are negligible.
As all three results are still similar, we allocate the moderate overall TCP goodput to the runtime
performance of the BMv2 P4 target.
The low throughput of \ac{BMv2} is due the fact that its use is intended for testing and not for production purposes.
Thus, the results show that the overhead of our IPsec implementation on BMv2 only slightly reduces the TCP goodput and the impact of encryption/decryption operations is negligible on this platform.

\subsection{Control Plane Evaluation}
We investigate how P4-IPsec's controller-based operation of IPsec affects the time needed for IPsec tunnel setup and renewal.

\subsubsection{Experiment Description}
In common IPsec deployment, \acp{SA} are set up between two IPsec peers using \ac{IKE} message exchange.
In P4-IPsec, an SDN controller sets up and renews IPsec tunnels, which may take longer due to controller operation and table updates on P4 switches.

For \textit{IPsec tunnel setup}, the SDN controller generates two unidirectional \acp{SA}, installs them on both P4 switches, and modifies the \ac{SPD} \acp{MAT} of both peers so that traffic is protected using \ac{IPsec}.
In our experiment, we measure the time for IPsec tunnel setup.
It starts when the southbound connections between the SDN controller and the two P4 switches are established and ends with the last confirmation of the MAT modifications on the two P4 switches.
For \textit{IPsec tunnel renewal}, the SDN controller generates one unidirectional \ac{SA} and installs it on both P4 switches.
Tunnel renewal is triggered by a P4 switch if the packet counter of a \ac{SA} reaches the soft packet limit.
The measurement is started on the controller when it receives the soft timeout notification from one P4 switch and it is stopped when the controller has received all confirmations of the P4 switches about all \ac{MAT} modifications.
Details of both operations including sequence diagrams can be found in \sect{tunnel-management}.
We recorded measurement data for IPsec tunnel setup and tunnel renewal within the experiment on TCP goodput for the AES\_CTR\_HMAC\_MD5 cipher suite as described before.

In the testbed environment, latency on the management link between the SDN controller and P4 switches is very low as all components run on the same host and as we have not configured any extra delay on the links.
In real-world deployments, link latencies are significant, but they impact both IKE message exchange and controller-based operation of IPsec.
By keeping the link latencies minimal, we derive an upper bound on potentially additional latency due to controller operation and MAT modification on P4 switches.

\subsubsection{Results \& Discussion}
\fig{setup-renewal-times} depicts the measured averages for IPsec tunnel setup and renewal times.
Tunnel setup takes \SI{5.02}{\milli\s} while the time for tunnel renewal is \SI{4.38}{\milli\s}.
These results show that control plane overhead is low.

To further investigate both operations, we we also analyze the durations of three major components: generating \ac{SA} data, inserting new \ac{MAT} entries, and updating existing {MAT} entries.
\fig{basic-operation-times} depicts their average times.
Generating keying material for a unidirectional \ac{SA} takes \SI{0.084}{\milli\s}.
Installing a new \ac{MAT} entry via a write operation takes \SI{0.702}{\milli\s}.
Updating an existing \ac{MAT} entry takes \SI{0.587}{\milli\s}.
Thus, the effort for key generation is almost negligible compared to \ac{MAT} modifications.
\section{Implementation on Hardware P4 Targets}
\label{sec:implementation-hardware}

In the following, we describe implementation experiences for the \sume and Edgecore Wedge 100BF-32X platform.

\subsection{NetFPGA SUME}
We give an overview of the platform and describe implementation experiences.

\subsubsection{Overview on Platform \& Development}
\sume is an open source hardware development board for prototyping network applications.
Its main part is a Xilinx Virtex-7 690T \ac{FPGA} with 4 SFP+ ports that acts as programmable data plane.
It supports throughput rates up to \SI{100}{\giga\Bit\per\s}.
The \sume board can be programmed via the \ac{SDNet} \cite{sdnet-px}, a proprietary predecessor of P4 from Xilinx.
Support for P4 programmability was introduced with the P4-NetFPGA tool \cite{p4-netfpga}.
First, a P4-to-SDNet compiler translates P4\textsubscript{16} programs into \ac{SDNet}.
Then, the \ac{SDNet} compiler generates \ac{HDL} blocks in Verilog that can be validated in generic and platform-specific \ac{FPGA} simulations.
Finally, the \ac{HDL} representation is synthesized into a hardware design to program the \ac{FPGA}.
In addition to the P4 program, custom functions can be implemented in a \ac{HDL} and included in the hardware design.
Programmers may implement custom \ac{HDL} blocks or integrate IP cores that can be used as P4 externs in the P4 program.
The P4-NetFPGA tool only supports the SimpleSumeSwitch architecture, i.e., P4 programs defined for more sophisticated architectures such as \ac{PSA} need to be transformed to this architecture.

\subsubsection{Implementation Experiences}
We report on implementation experiences about porting our software-based implementation P4-IPsec for the \sume board.
First, P4-NetFPGA is currently limited to \textit{packet header manipulation}.
P4-IPsec requires modifications of packet payloads, i.e., we were required to parse packet payloads as an additional header field.
As P4-NetFPGA does not support parsing variable-length header fields, the implementation is limited to packets with a fixed length.
Second, P4-NetFPGA lacks a packet streaming function for \textit{data exchange between the P4 pipeline and P4 externs}.
Instead, data between the P4 pipeline and externs is currently exchanged via blocks of bits.
As this data transfer needs to be executed within one clock cycle of the \ac{FPGA}, the data size is limited.
We observed that this limit is about \SI{10}{\kilo\bit} for one function call.
This limits the maximum packet size to be processed through a P4 extern to about \SI{600}{\byte}.
During the synthesis, the Vivado suite optimizes the hardware implementation through several algorithms.
In various experiments, we observed a practical upper bound of about \SI{140}{\byte} for packets.
Either the hardware implementation did use more resources than offered by the \ac{FPGA}, or data transfer and calculation within the P4 extern exceeded one clock cycle.
A packet streaming function was announced in 2018, but is still not available.
Last, we encountered several \textit{more general problems} with P4-SDNet and the \sume board.
Probably due to a bug, we were not able to access the values of an LPM table for IP routing with our \ac{SDN} controller.
We solved that problem by using exact matching tables instead, an approach that is not acceptable for a production implementation.
In addition, we experienced several stability problems.
No matches in \acp{MAT} were found when data was written to hardware registers.
Finally, we missed many important details in the documentation.
With hope for improved support, we managed to implement a very limited prototype.
It only allows to apply the NULL cipher on fixed-length packets that do not exceed a total length of \SI{140}{\byte}.

Scholz et al. \cite{ScOe19} report on implementation experiences of cryptographic hashing functions in P4 data planes.
The \sume board is also one of the platforms examined where the authors present results that correspond to our results.
As a workaround, the authors propose to move the externs subsequent to the synthesized P4 program.
However, the workaround can be applied only if the P4 program does not rely on the output of the extern.
This makes it inapplicable to P4-IPsec.
Besides, implementing this workaround requires extensive knowledge about \acp{HDL} and \ac{FPGA} programming.

\subsection{Edgecore Wedge with Tofino}
We give an overview of the platform and describe why a direct adoption of P4-IPsec is not feasible.
We present two workaround implementations and evaluate their performance in experiments.

\subsubsection{Overview on Platform \& Development}
Our testbed switch, the Edgecore Wedge 100BF-32X \cite{wedge-datasheet}, features 32 QSFP28 network ports with throughput rates up to \SI{100}{\giga\bit\per\s}.
The QSFP28 ports interface the Tofino switching ASIC from Barefoot Networks which is fully programmable with P4.
The Tofino ASIC connects to a CPU module via PCIe.
It features an Intel Pentium D1517 processor (\SI{1.6}{\giga\hertz}, 4 cores), \SI{8}{\giga\byte} RAM, and a \SI{32}{\giga\byte} SSD.
For programming and managing the Tofino ASIC, the CPU module runs the Barefoot P4 Software Development Environment on top of a Linux-based operating system.
It loads and manages P4 programs during execution, provides management APIs (e.g., P4Runtime), and exposes an interface for network packet exchange between the P4 processing pipeline and the CPU module.
Due to its optimization for high-speed packet processing with bandwidths up to multiple \si{\tera\bit\per\s} in data center or core networks, user-defined P4 externs that may contain computation-intense functions are not supported.

\subsubsection{Workaround Implementations}
\label{sec:wedge-implementation}

\begin{figure}[ht]
    \begin{center}
    \includegraphics[width=.99\linewidth]{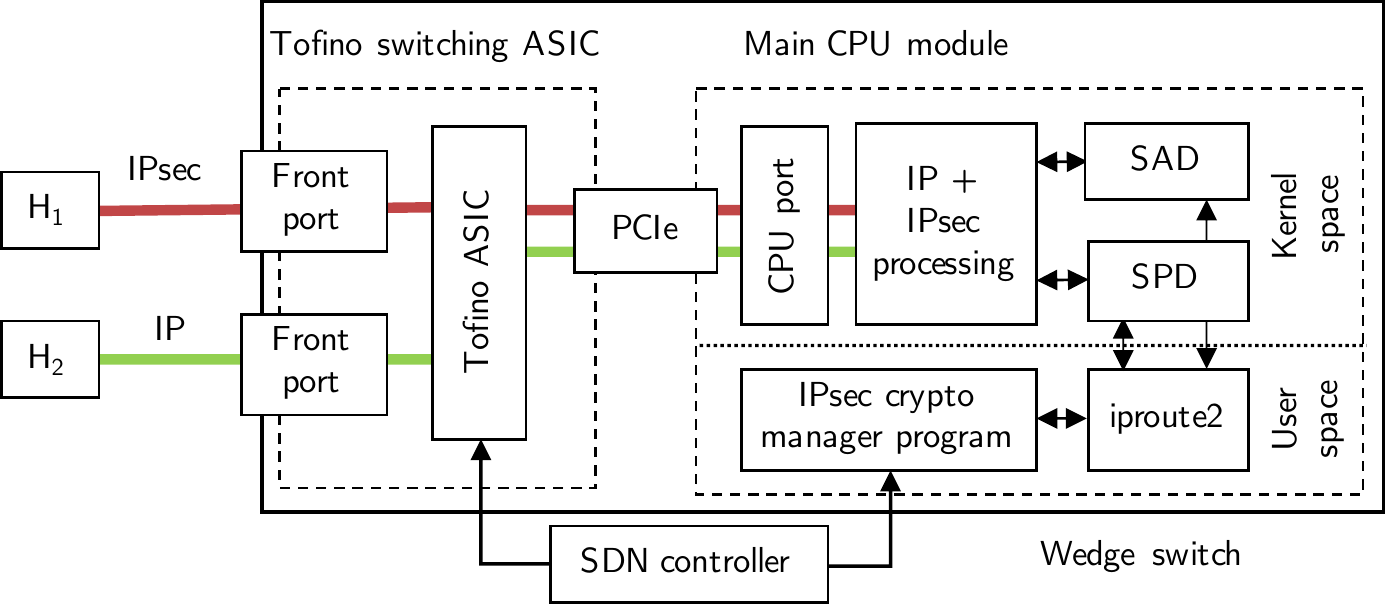}
    \end{center}
    \caption{First workaround implementation. We relocate \ac{IPsec} processing to the main CPU module that interfaces the Tofino switching \ac{ASIC} via a PCIe CPU port.}
    \label{fig:tofino-1}
\end{figure}

In our first workaround implementation, we \textit{relocate the P4 externs of P4-IPsec to the main CPU module}.
\fig{tofino-1} depicts the concept.
We replace all P4 extern function calls in the P4 processing pipeline by packet transfers via the CPU port to the main CPU module.
On the main CPU module, we use the IPsec kernel functions of the Linux operating system for IPsec processing.
We implement a simple IPsec crypto manager program that translates P4-IPsec configuration from the controller into IPsec configuration for the Linux host.
We implement the IPsec crypto manager in Python 3.
It relies on iproute2 commands to manage the \ac{SPD} and \ac{SAD} of the Linux host.

We briefly evaluate this first workaround implementation with experiments on latency and TCP goodput.
As depicted in \fig{tofino-1}, we attach two physical hosts running Ubuntu 16.04 LTS via \SI{100}{\giga\bit\per\s} links to the front ports of the Wedge switch.
We enable IPsec on the link between H\textsubscript{1} and the switch while the link between the switch and H\textsubscript{2} remains unprotected.
First, we measure the latency that is introduced by IPsec processing and packet exchange with the CPU module.
We send \num{100} ICMP echo requests from H\textsubscript{1} to H\textsubscript{2} and measure an average round-trip time of about \SI{1.5}{\milli\s}.
Second, we investigate on the maximum TCP goodput.
We generate TCP transmissions with iperf3 in three experiments, each performed with five runs and a duration of \SI{30}{\s}.
For getting a reference, we measure the maximum TCP goodput between the P4 processing pipeline and the main CPU module.
Therefore, we assign an IP address to the virtual network interface of the CPU port on the main CPU module and run iperf3 measurements between H\textsubscript{1} and the main CPU module.
We measure an average TCP goodput of about \SI{3.3}{\giga\bit\per\s}.
We consider this as upper bound for the main CPU module.
Now, we measure TCP goodput between H\textsubscript{1} and H\textsubscript{2}.
When using the NULL cipher suite, we measure an average TCP goodput of about \SI{2}{\giga\bit\per\s}.
For the AES-GCM-256 cipher suite, the average TCP goodput drops to about \SI{1.4}{\giga\bit\per\s}.
We repeat the experiment for \num{16} concurrent IPsec tunnels and calculate the average of \num{10} runs with a duration of \SI{300}{\s}.
The maximum TCP goodput remains at \SI{2}{\giga\bit\per\s} for IPsec with the NULL cipher suite and \SI{1.4}{\giga\bit\per\s} for IPsec with the AES-GCM-256 cipher suite.
We attribute the large differences in TCP goodput to the rather slow CPU with a base frequency of \SI{1.6}{\giga\hertz}.
Still, we consider this a very reasonable performance that might be sufficient for scenarios where only few shared network resources should be accessed sporadically by roadwarrior hosts.

\begin{figure}[ht]
    \begin{center}
    \includegraphics[width=.94\linewidth]{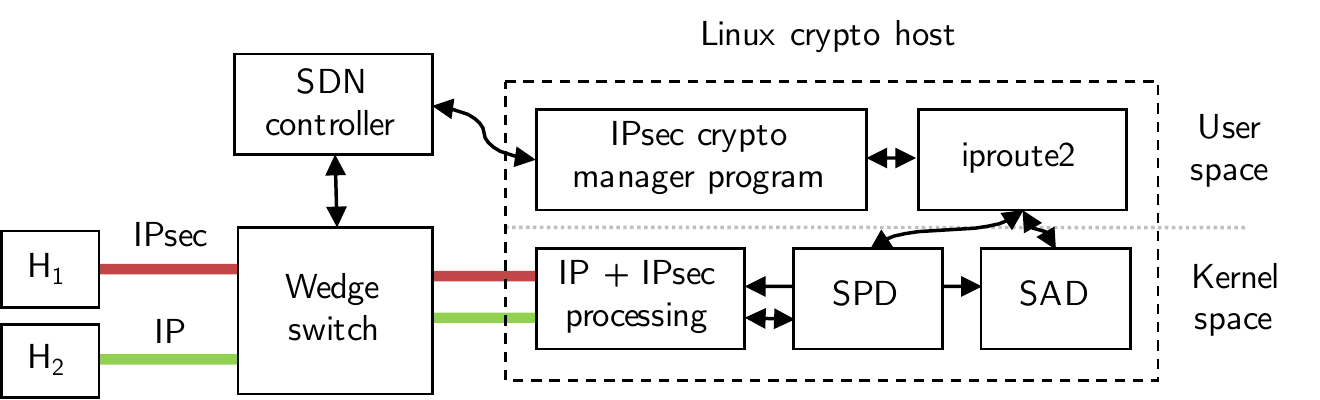}
    \end{center}
    \caption{Second workaround implementation. We forward IPsec-related flows to a Linux crypto host.}
    \label{fig:tofino-3}
\end{figure}

In our second workaround implementation, we \textit{forward IPsec-related flows to a crypto host}.
This approach is used by several past works on integrating IPsec with fixed-function SDN data planes (e.g., \cite{VaKa17}).
As depicted in \fig{tofino-3}, we set up a Linux crypto host for offloading IPsec processing.
We deploy the IPsec crypto manager program from the previous workaround implementation as interface between the crypto host and controller.
We deploy a simple P4 program on the Wedge switch that forwards IPsec flows based on a \ac{MAT}.
The controller writes/edits the forwarding \ac{MAT} on the Wedge switch and sends configuration messages to the IPsec crypto manager program.

\begin{figure}[ht]
    \begin{center}
    \includegraphics[width=.94\linewidth]{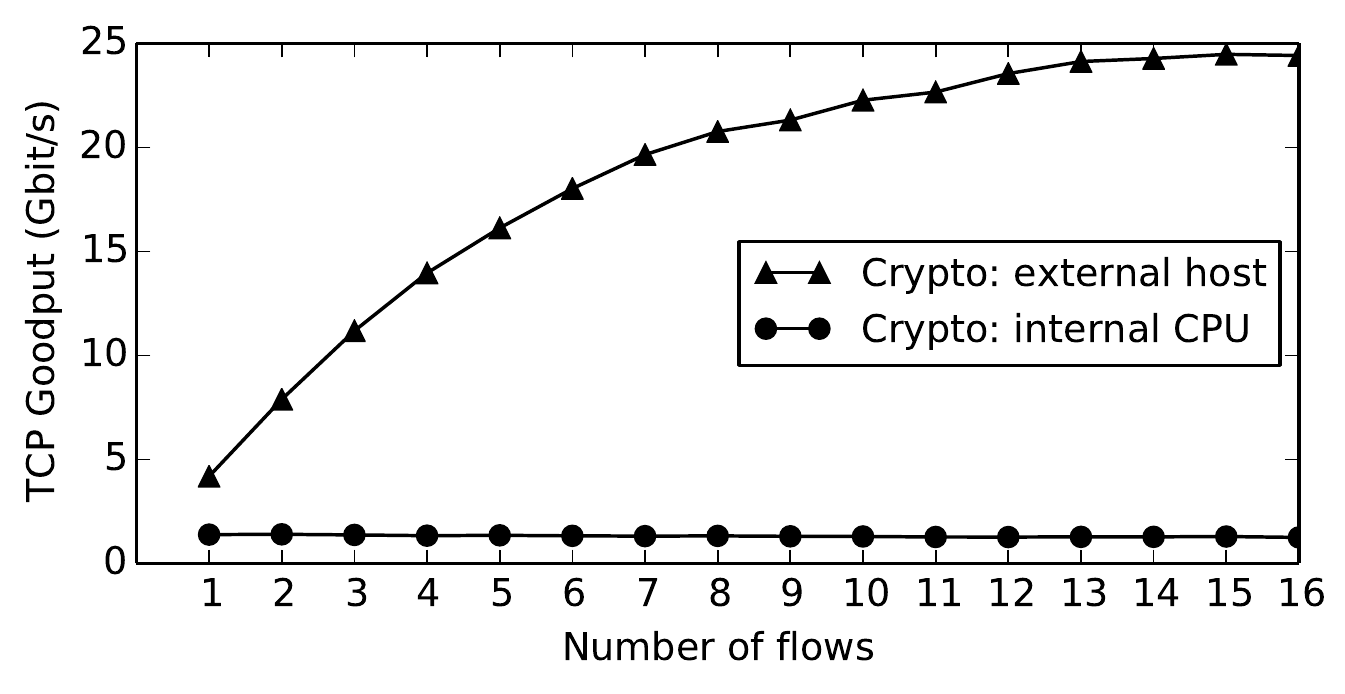}
    \end{center}
    \caption{Average TCP goodput for both workaround implementations and 1-16 IPsec tunnels with the AES-GCM-256 cipher suite.}
    \label{fig:throughput-tofino}
\end{figure}

For a simple evaluation, we set up a crypto host with an Intel Xeon Gold 6134 CPU (8 cores, 16 threads), \SI{128}{\giga\byte} RAM, and a \SI{240}{\giga\byte} SSD, running Ubuntu 18.04 LTS.
We perform the same experiments as for the first workaround implementation.
The round-trip time is about \SI{2}{\milli\s} which is slightly larger than in the previous approach.
\fig{throughput-tofino} compares TCP goodput results for IPsec tunnels with the AES-GCM-256 cipher suite of both workaround implementations.
For a single IPsec tunnel with the AES-GCM-256 cipher suite, we measure an average TCP goodput of about \SI{4}{\giga\bit\per\s}.
It can be increased by running multiple connections over the same crypto host.
For 16 parallel IPsec tunnels, we measure an overall average TCP goodput of about \SI{24}{\giga\bit\per\s}.
This effect can be attributed to receive-side scaling (RSS) of the network interface card, which can leverage multiple cores, but only one per IPsec tunnel. 
In case of multiple IPsec tunnels, the overall TCP goodput can be increased through RSS by leveraging the  processing power of more than a single core.
Crypto capacity can be scaled up by increasing the number of crypto hosts connected to the switch.
TCP goodput on each crypto host can be further improved by optimization techniques as presented in \sect{ipsec-packet-processing}.

Chen \cite{Ch20} presents an implementation of AES encryption for Tofino-based P4 switches.
It uses a novel Scrambled Lookup Table technique that allows througput rates of up to \SI{10.92}{\giga\bit\per\s} for AES-128.
However, the current concept is limited to blockwise encryption of packets with a maximum payload size of \num{16}~bytes so that it is in its current form not a suitable base for IPsec support.
If subsequent versions of this work introduce block chaining, integrating P4-IPsec could be an interesting follow-up work.
\section{Conclusion}
\label{sec:conclusion}

In this work, we proposed the first implementation of IPsec in P4.
The proposed data plane implementation features ESP in tunnel mode and provides support for multiple cipher suites with the help of P4 externs.
P4-IPsec supports automated operation of IPsec in host-to-site and site-to-site scenarios.
IPsec tunnels are set up and managed by an SDN controller based on predefined tunnel profiles.
For interaction with remote hosts in host-to-site scenario, we introduce a client agent for Linux hosts.
We introduced the fundamentals of IPsec and data plane programming with P4, gave an extensive review on related work, and presented the architecture of P4-IPsec.

P4 programmable data planes open up the possibility of implementing IPsec on SDN-capable data plans for the first time.
However, the implementation on P4 switches is still challenging.
For the \ac{BMv2} software switch, the implementation was straightforward, but moderate data rates make its practical application difficult.
The controller-supported signaling was not a bottleneck, however.
Due to platform limitations of the NetFPGA SUME board, we were not able to build a working prototype.
Our results for the Tofino-based Wedge switch are more promision:
Even though this P4 switch does not support P4 Externs, we presented two workaround implementations that either use the main CPU module or an external crypto host.

We have shown that security use cases can benefit from P4, but crypto functions are still missing on many P4 hardware switches.
Therefore, we advocate for P4 hardware targets that either include P4 externs for those operations or offer powerful interfaces so that developers can run individual functions on the CPU module of such switches.
Such features have the potential to massively foster the deployment of P4 targets in practice and stimulate further network research.

\section*{List of Acronyms}
\vspace{0.2cm}
\begin{acronym}[SMGW-PP]

\acro{SDN}{software-defined networking}
\acro{ONF}{Open Network Foundation}
\acro{OF}{OpenFlow}
\acro{ODP}{Open Data Plane}
\acro{NFV}{Network Function Virtualization}
\acro{VNF}{virtual network function}
\acro{SFC}{Service Function Chaining}
\acro{BMv2}{Behavioral Model version 2}
\acro{MAT}{match-and-action table}

\acro{VPN}{Virtual Private Network}
\acro{IP}{Internet Protocol}
\acro{IPsec}{Internet Protocol Security}
\acro{ESP}{Encrypted Secured Payload}
\acro{AH}{Authentication Header}
\acro{PPF}{packet processing function}
\acro{SP}{security policy}
\acro{SPD}{Security Policy Database}
\acro{SA}{security association}
\acro{SAD}{Security Association Database}
\acro{PAD}{Peer Authentication Database}
\acro{SPI}{Security Parameter Index}
\acro{IKE}{Internet Key Exchange}
\acro{IKEv2}{Internet Key Exchange v2}
\acro{IPComp}{IP Payload Compression}

\acro{AE}{authenticated encryption}
\acro{ICV}{Integrity Check Value}
\acro{IV}{initialization vector}
\acro{3DES}{Triple Data Encryption Standard}
\acro{AES}{Advanced Encryption Standard}
\acro{CBC}{cipher block chaining}
\acro{CTR}{counter}
\acro{GCM}{galois/counter mode}
\acro{HMAC}{keyed-hash message authentication code}
\acro{SHA}{secure hash algorithm}
\acro{TLS}{Transport Layer Security}
\acro{PKI}{public key infrastructure}

\acro{SoC}{system on a chip}
\acro{FPGA}{field programmable gate array}
\acro{NPU}{network processing units}
\acro{ASIC}{application-specific integrated circuit}
\acro{NIC}{network interface card}
\acro{NPU}{network processing unit}
\acro{APU}{accellerated processing unit}
\acro{DPDK}{Data Plane Development Kit}
\acro{SDNet}{Software Defined Specification Environment for Networking}
\acro{HDL}{hardware description language}
\acro{HLIR}{high-level intermediate representation}
\acro{RTL}{Register Transfer Level}
\acro{PSA}{Portable Switch Architecture}

\acro{TTL}{time to live}
\acro{INT}{in-band network telemetry}
\acro{LPM}{longest-prefix matching}
\acro{CLI}{command line interface}
\acro{FSM}{finite state machine}
\acro{API}{application programming interface}
\acro{LLDP}{Link Layer Discovery Protocol}
\acro{MACsec}{Media Access Control Security}

\setlength{\parskip}{0ex}
\setlength{\itemsep}{0ex}

\end{acronym}

\section{Acknowledgement}
The authors would like to thank the anonymous reviewers for their valuable comments.

\bibliography{literature}
\bibliographystyle{unsrt}

  
\end{document}